\documentclass[%
 reprint,
 amsmath,amssymb,
 aps,
]{revtex4-2}
\usepackage[utf8]{inputenc}
\usepackage{graphicx}% Include figure files
\usepackage{dcolumn}% Align table columns on decimal point
\usepackage{bm}% bold math
\usepackage{physics}
\usepackage{xcolor}
\usepackage{lipsum}

\begin{document}

\preprint{APS/123-QED}

\title{Enhancing strontium clock atom interferometry using quantum optimal control}
\author{Zilin Chen, Garrett Louie, Yiping Wang, Tejas Deshpande, Tim Kovachy}
\affiliation{%
 Department of Physics and Astronomy and Center for Fundamental Physics, Northwestern University}%

\begin{abstract} 
 Strontium clock atom interferometry is a promising new technology, with multiple experiments under development around the world to explore its potential for dark matter and gravitational wave detection. In these detectors, large momentum transfer (LMT) using sequences of many laser pulses is necessary, and thus high fidelity pulses are important since small errors become magnified. Quantum Optimal Control (QOC) is a framework for developing control pulse waveforms that achieve high fidelity and are robust against experimental imperfections. Resonant single-photon transitions using the narrow clock transition of strontium involve significantly different quantum dynamics than more established atom interferometry methods based on far-detuned two-photon Raman or Bragg transitions, which leads to new opportunities and challenges when applying QOC.  Here, we study in simulation QOC pulses for strontium clock interferometry and demonstrate their advantage over basic square pulses (primitive pulses) and composite pulses in terms of robustness against multiple noise channels.  This could improve the scale of large momentum transfer in Sr clock interferometers, paving the way to achieve these scientific goals.
\end{abstract}

\maketitle
\section{Introduction}
Light-pulse atom interferometry is a technology that uses laser pulses to split, manipulate, and recombine the motional states of atoms so that precise measurements can be made from their interference. It has proven itself a powerful tool for precision metrology and sensing, with applications including tests of quantum mechanics and the equivalence principle \cite{Zhou2015_equivalence, rosi2017equivalence, Overstreet2018_equivalence, Asenbaum2017,Asenbaum2020_equivalence, kovachy2015_halfmeter, fray2004atomic, Schlippert2014ep, Barrett2016, kuhn2014bose, barrett2015correlative, PhysRevLett.113.023005, PhysRevA.88.043615, Hartwig2015, williams2016quantum, overstreet2022observation}, terrestrial and spaceborne gravitational wave detection \cite{Dimopoulos2008_GW, graham2013new, Graham2016_GW, PhysRevD.93.021101, canuel2018exploring, hogan2011atomic, abou2020aedge, Badurina_2020, ZAIGA2020, abe2021matter}, precision measurements of the fine structure constant \cite{Bouchendira2011_finestructure, parker2018finestructure, morel2020determination} and gravity \cite{Biedermann2015_gravity, rosi2014precision}, searches for dark matter \cite{Arvanitaki2018_darkmatter, Graham2016_darkmatter,banerjee2022phenomenology} and dark energy \cite{hamilton2015_darkenergy}, and mobile surveying \cite{Wu2019_mobile, bongs2019taking}. The most sensitive interferometers employ large momentum transfer (LMT) techniques, which increase the enclosed spacetime area with additional laser pulses. Typical LMT atom optics utilize far-detuned, multi-photon transitions \cite{mcguirk2000, muller2008, muller2009, clade2009, gebbe2021twin, 102hk_large_area, Close:2013, Mazzoni2015, Kotru2015, Plotkin2018, pagel2019bloch}, where two ground states are coupled via a short-lived excited state. The scaling of such pulse sequences is limited by spontaneous emission, which can only be mitigated so long as additional laser power is available. STIRAP has been used in atom interferometry but does not currently match the performance of other state-of-the-art atom optics, limited by other detuned excited states \cite{young1997measurement}. 

In contrast, alkaline earth atoms such as Sr possess long-lived excited states, which have been leveraged to achieve state-of-the-art atomic clocks \cite{hinkley2013clock, bloom2014clock}. In $^{87}{\rm Sr}$, the $^1{\rm S}_0 \to {}^3{\rm P}_0$ clock transition has a lifetime over 100 seconds, allowing resonant, single-photon atom optics with greatly reduced spontaneous emission losses. This affords an enormous increase in available pulse area, potentially scalable to thousands of pulses before significant spontaneous emission losses \cite{rudolf2020_689LMT, abe2021matter}. 
 
A clock interferometer is a specific type of atom interferometer using such resonant, single-photon transitions. It offers improved laser phase noise rejection in differential measurement configurations comparing multiple interferometers over a long baseline \cite{Graham2016_GW} \footnote{for atom interferometers based on two-photon atom optics, multiple baselines can be used to achieve improved laser noise suppression \cite{canuel2018exploring}}, which is essential for dark matter and gravitational wave detection. Moreover, the clock transitions in alkaline earth atoms are orders of magnitude less susceptible to magnetic fields than in the alkalis \cite{intercombination_line_magnetig_field}. This new generation of recently demonstrated single-photon clock atom interferometers \cite{Hu2017_clockinterferometry, hu2019sr, rudolf2020_689LMT, wilkasonFloquet} is poised to study hitherto elusive phenomena such as ultralight, wavelike dark matter  \cite{Arvanitaki2018_darkmatter, Graham2016_darkmatter, abe2021matter, ZAIGA2020, Badurina_2020, abou2020aedge,banerjee2022phenomenology}, tests of atom charge neutrality \cite{Arvanitaki2008_neutrality}, mid-band gravitational wave detection \cite{abe2021matter,Badurina_2020,abou2020aedge, ZAIGA2020}, and tests of quantum mechanics at unprecedented delocalization scales \cite{abe2021matter}.

In practice, the performance of atom interferometers is also limited by the noise in the driving field and inhomogeneities across the atom cloud. Atom losses and phase errors caused by these effects accumulate with repeated pulses and thus limit the scaling of LMT systems. Simple robust control pulses for two-level quantum systems under detuning and amplitude errors were developed for nuclear magnetic resonance (NMR) spectroscopy \cite{levitt1986composite,  emsley1992optimization,NMRComputation} and have more recently found popularity in quantum computation \cite{Cummins2003compositecompute, Collin2004compute}. More sophisticated pulse shaping algorithms, such as gradient ascent pulse engineering (GRAPE) \cite{khaneja2005optimal} and chopped random basis (CRAB) optimization \cite{Doria2011_CRAB, Caneva2011_CRAB}, minimize a cost function that quantifies the infidelity of the operation. As this may be an arbitrary function of hundreds of parameters, highly modulated control pulses can now be tailor-made for driving particular quantum dynamics with an appropriate cost metric. For instance, quantum optimal control (QOC) is a powerful tool for realizing high-fidelity gates in quantum computing \cite{mueller2022one, grace2007optimal, Rebentrost2009optimal, Abdelhafez2020optimal,koch2022quantum} and levitated nanoparticle control \cite{magrini2021real,weiss2021large}.

In the field of atom interferometry, quantum control schemes including composite pulses \cite{Butts2013, dunning2014_composite, Berg2015_composite}, shaped pulses \cite{Luo2016_shaped_pulses}, adiabatic rapid passage \cite{Kotru2015, tim_ARP_Bragg}, and numerical optimal control \cite{saywell2018_QuOC_Mirror, saywell2020optimal, saywell_2020_biselective, goerz2021quantum, Goerz2023} have been applied to Raman and Bragg transitions with alkalis, and Floquet pulse engineering has been applied to single-photon atom optics on the $^1{\rm S}_0 \to {}^3{\rm P}_1$ transition of $^{88}{\rm Sr}$ \cite{wilkasonFloquet}. In this paper, we report simulation studies of numerical optimal control's applications to $^1{\rm S}_0 \to {}^3{\rm P}_0$ clock interferometry with $^{87}{\rm Sr}$. Using the gradient-based optimization tools in Q-CTRL's BOULDER OPAL package \cite{ball2021software}, we design optimized analogs of $\pi$ pulses which are robust to as many as five simultaneous noise channels. Sequences of many of these pulses can be used to achieve enhanced LMT \cite{graham2013new}.  Under an experimentally relevant range of noise, we find that the optimized pulses have fidelity improved beyond that of basic square (primitive) pulses by an order of magnitude or more. Successful implementation of these pulses may allow atom interferometers to overcome unavoidable limitations and realize their full potential.

A challenge in clock interferometry is that imperfections during laser beam delivery or local changes to the quantization axis from stray magnetic fields can introduce undesired polarization components, which couple transitions to additional magnetic sublevels (Fig. \ref{fig:setup}(a)). Larger bias fields can suppress errors in the quantization axis, but are unfavorable due to quadratic Zeeman shifts \cite{abe2021matter} and cannot mitigate errors already present in the beam. It is therefore a valuable application of quantum optimal control to design pulses that are insensitive to polarization defects, reducing the atom loss and phase errors arising from the population of other sublevels.

\section{Optimization Protocol}

\begin{figure}[ht]
    \centering
    \includegraphics[width=\linewidth]{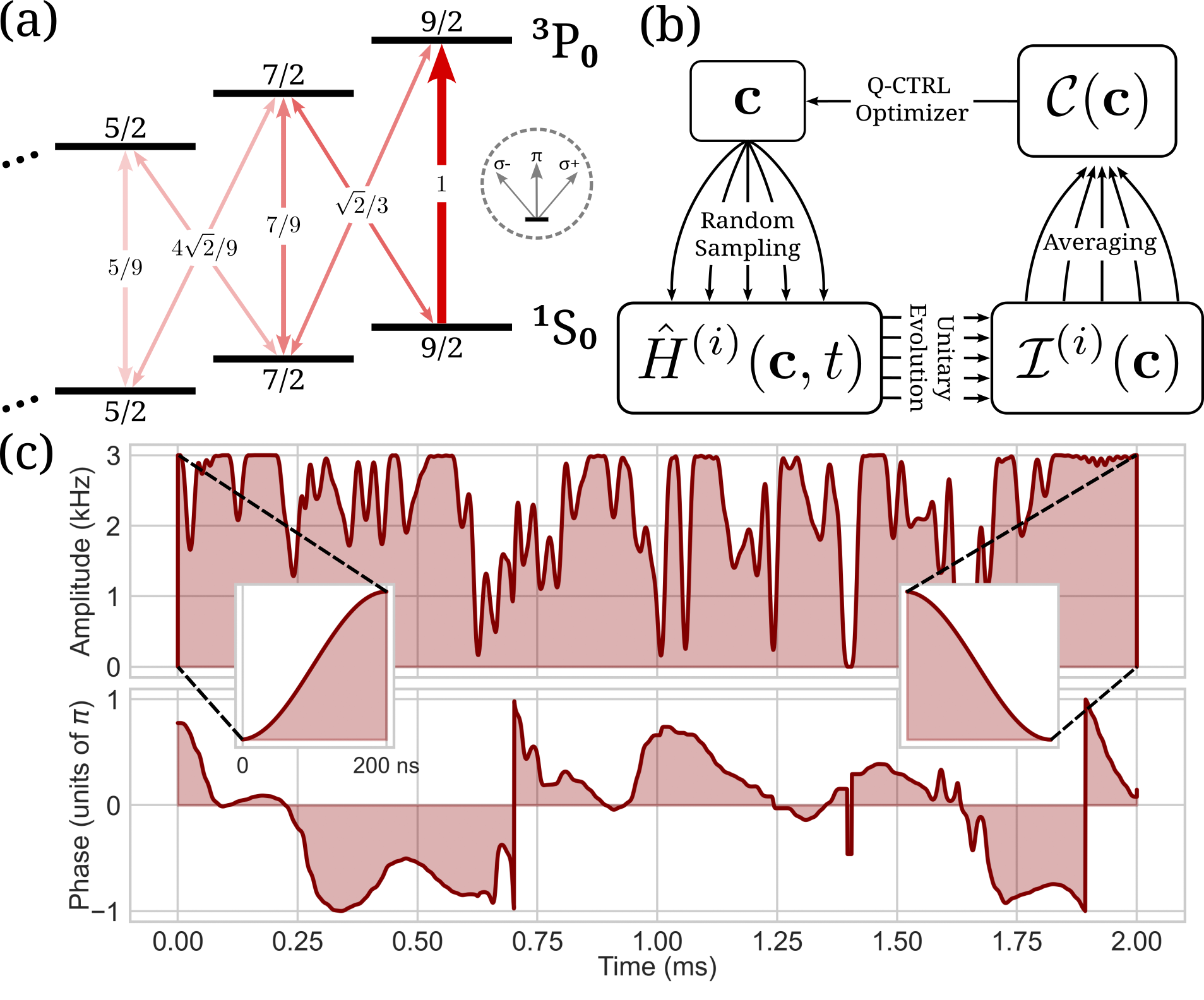}
    \caption{(\textbf{a}) Substructure of the $^1{\rm S}_0 $ and $ ^3{\rm P}_0$ manifolds, with transition strengths indicated. Parasitic $\sigma_+$ and $\sigma_-$ polarizations 
    couple to states outside the desired $m_F=9/2\to m_F'=9/2$ two-level system. (\textbf{b}) Optimization procedure. Randomly sampled noise values for multiple noise channels define a batch of noise trajectories. The unitary evolution of the system is calculated under each, yielding a batch of infidelities which are averaged to give the final cost function. The scope of robustness is tuned by modifying the sampling distributions. (\textbf{c}) The optimized pulse starts and ends with a smooth switching (shown in the insets). The phase and amplitude modulation are smoothed with a Gaussian-weighted moving average filter. The frequency components higher than $4\times10^4$ Hz are filtered out. The frequency spectrum is within the modulation bandwidth of acousto-optic modulators (AOMs), and the switching on and off process is within the rise time achievable by AOMs. The smoothed pulse is used for all simulations.}
    \label{fig:setup}
\end{figure}

In the interferometry scheme studied here, a cloud of $^{87}{\rm Sr}$ atoms is initialized in the $\ket{^1{\rm S}_0;m_F=9/2}$ state and driven to the $ \ket{^3{\rm P}_0; m=9/2}$ state on the 698 nm clock transition, which has a natural linewidth of 1 mHz \cite{abe2021matter}. For optimization, we split a pulse of fixed duration uniformly into $N$ time segments, whose amplitude and phase become the $2N$ independent variables\textemdash hereafter referred to as $\textbf{c}$\textemdash which are modified during optimization. The pulse duration and the number of segments are chosen before optimization, as is the learning rate (step size in the gradient-based search). We vary these parameters between optimization runs to determine what gives the best results (Fig. \ref{fig:setup}(b)). We choose maximum Rabi frequencies of several kHz \cite{abe2021matter} and find pulse lengths of 2 ms or longer best for optimizing against all noise channels. With the many-second interferometer duration in tall atomic fountains \cite{abe2021matter} or spaceborne detectors \cite{Dimopoulos2008_GW, hogan2011atomic, abou2020aedge}, sequences of thousands of pulses may be performed, further increased by future upgrades to laser power \cite{bertoldi2021fast, nourshargh2021circulating}.  In the optimized pulse, the segment time is approximately 17 $\mu s$, which is significantly longer than the response time of an acousto-optic modulator (AOM).

\begin{figure*}
    \centering
    \includegraphics[width=\textwidth]{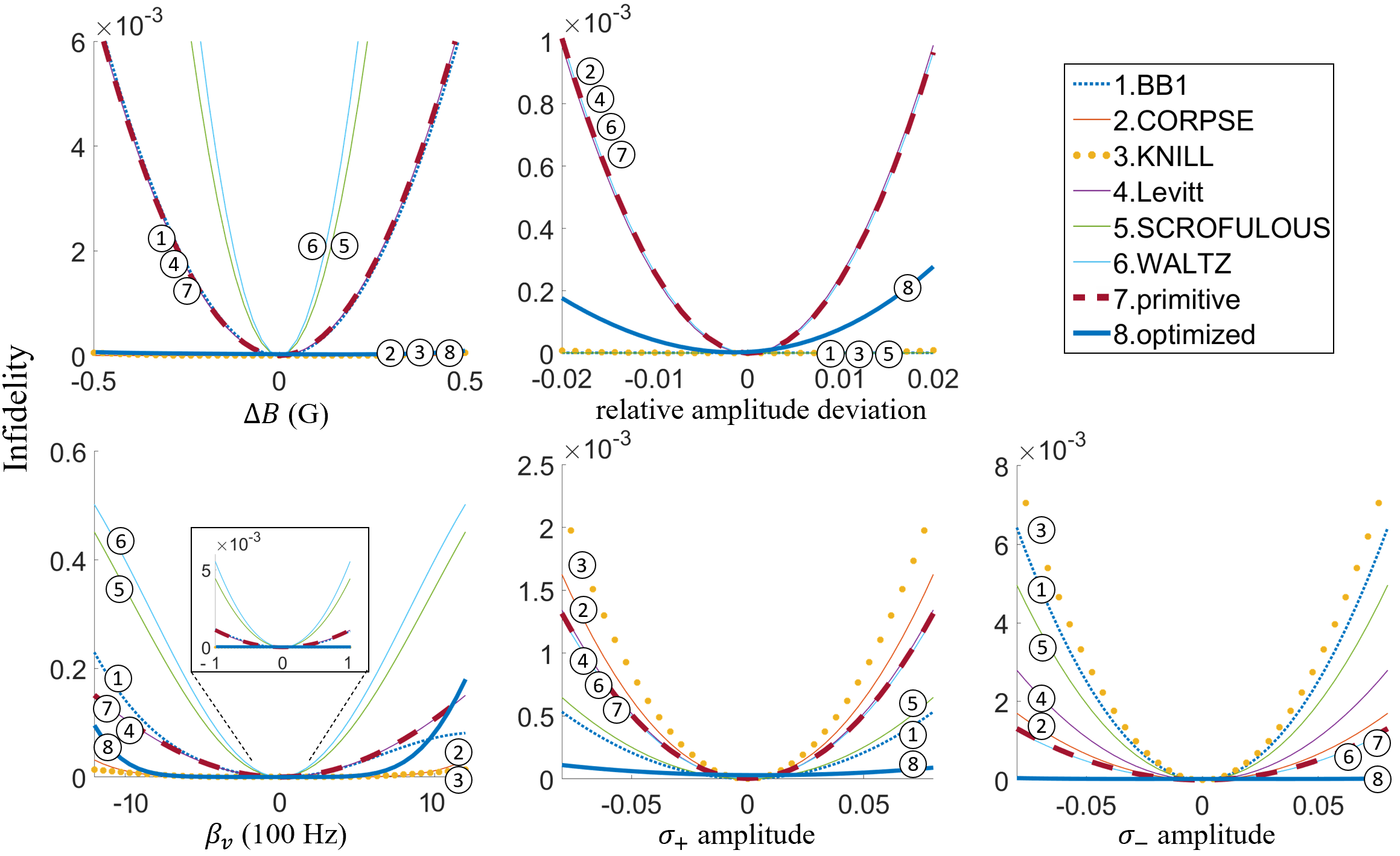}
    \caption{Comparison of the optimized pulse to primitive pulse and various composite pulses \cite{saywell2018_QuOC_Mirror,levitt1986composite}. Single pulse infidelity with noise parameters' variation are compared among the optimized pulse, primitive pulse, and composite pulses. $\Delta B$ is the deviation of the bias magnetic field from its nominal value $B_0 = 1$ G, so that the total bias field is $B = B_0 + \Delta B$. $\beta_v$ is the detuning due to the Doppler shift in units of 100 Hz. The units of the horizontal axes for the polarization and  amplitude plots correspond to fractional errors. In contrast to all other pulses, the optimized pulse maintains a high degree of robustness against all noise channels and outperforms all other pulses in robustness against polarization noise.}
    \label{fig:lineprofile}
\end{figure*}

To achieve robust control, we use a cost function based on random samplings of noise trajectories (one trajectory is the Hamiltonian with a particular set of noise parameters) \cite{goerz2014robustness}. We include five channels of static noise: $\sigma_{+}$ and $\sigma_{-}$ polarization components, which couple sublevels outside the desired two-level system; amplitude noise on the drive; detuning errors from Doppler shifts in the atom cloud; and variation in the bias field, which changes the Zeeman splitting of the sublevels. Each noise channel is associated with a Hamiltonian term (see Appendix A). For each of these operators, we randomly sample $N_B$ sets of noise amplitudes\textemdash denoted as $\boldsymbol\beta^{(i)} = (\epsilon_+^{(i)}, \epsilon_-^{(i)},\beta_A^{(i)}, \beta_{v}^{(i)} ,\beta_B^{(i)})$ for the $i$th noise trajectory in the batch\textemdash from Gaussian distributions whose predefined width determines the scope of desired robustness. 

For each of these Hamiltonians $\hat{H}^{(i)}(\textbf{c}, t)$, we calculate the unitary evolution, from which we determine an infidelity $\mathcal{I}^{(i)}(\textbf{c})$. The infidelity is defined as
\begin{equation}
    \mathcal{I}^{(i)}(\textbf{c})=1-\left|\frac{\operatorname{Tr}\left(U_{\text {target }}^{\dagger} U^{(i)}(\textbf{c})\right)}{\operatorname{Tr}\left(U_{\text {target }}^{\dagger} U_{\text {target }}\right)}\right|^{2},
    \label{eqn:gateinfidelity}
\end{equation}
where $U_{\text {target }}$ is the target state evolution, and $U^{(i)}$ is the evolution due to the Hamiltonian under the $i$th noise trajectory \cite{ball2021software}. This metric is sensitive to both the population transfer and the phase imprinted by the pulse, which is important for atom interferometry.
All 20 ground and excited sublevels are included in our simulations. To construct the total cost function, we average the infidelities from all trajectories:
\begin{equation}
    \mathcal{C}(\textbf{c}) = \frac{1}{N_B} \sum\limits_{i=1}^{N_B}\mathcal{I}^{(i)}(\textbf{c}),
\end{equation}
where we typically use a batch size $N_B$ of 200. We found that increasing to a larger batch size did not provide significant improvement. The gradient-based optimizer in BOULDER OPAL determines the control variables $\textbf{c}$ which minimize this cost function. Thus, the pulse waveforms (Fig. \ref{fig:setup}(c)) are determined which retain low infidelity even in the presence of noise. 

Alternative methods of optimizing composite pulses for multilevel systems have also been previously investigated by other groups. Caneva et al. \cite{Caneva2011_CRAB} investigated the efficiency of the chopped random basis (CRAB) technique in optimizing different quantum processes. Genov and Vitanov \cite{genov2013dynamical} also used composite pulses to achieve efficient population transfer in multi-state quantum systems. Compared to this study, our QOC method, instead of analyzing the system's mathematical process in a Taylor expansion approximation to determine the optimal pulses, numerically searches in a large parameter space to find the most robust pulse shape. In the future work, it will be interesting to compare the results of different optimization methods applied to clock atom interferometers. 

\section{Results and Comparison}
\subsection{Single optimized pulse}
\begin{figure}
    \centering
    \includegraphics[width=\linewidth]{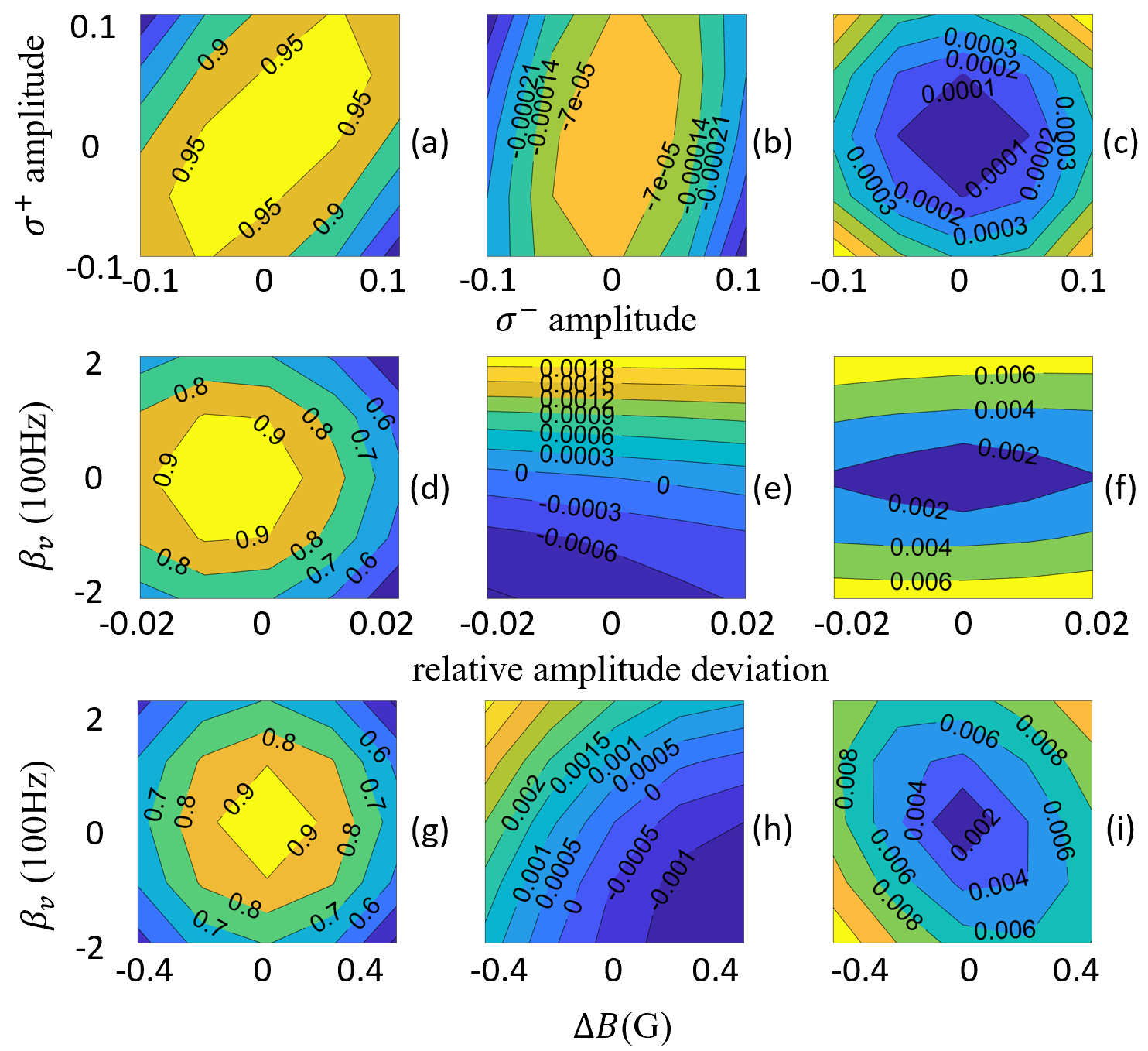}
    \caption{Full interferometer simulation. The results from the optimized pulse are presented. The time-dependent noise level is the same as in Fig. 3. (a), (d), (g): the transfer efficiencies (including the net effect of all pulses) with static noise on polarizations, Doppler detuning, magnetic field deviation, and amplitude. (b), (e), (h): the mean phase difference (rad) between two arms in the interferometer. (c), (f), (i): the standard deviation of the phase difference (rad) with random noise (half percent RMS in the amplitude and one percent RMS in the phase).}
    \label{fig:interferometer}
\end{figure}

To visualize the performance of various pulse schemes, we scan the infidelity across values of one (Fig. \ref{fig:lineprofile}) or two (Fig. \ref{fig:interferometer}) noise channels. In Fig. \ref{fig:lineprofile}, the optimized pulse performs up to an order of magnitude (or more) better than the primitive and composite pulses for the range of polarization errors considered here while maintaining strong robustness against amplitude and detuning errors. The range of noise values in optimizations and these plots follow what is expected in relevant experiments. In dark matter and gravitational wave experiments, it is favorable for the atom cloud's spatial extent to remain within the central region of the laser beam to help mitigate systematic errors such as those due to laser wavefront perturbations and residual AC Stark shifts from far-detuned transitions \cite{abe2021matter}. Therefore the laser amplitude variation across the atom cloud is assumed to be within several percent. We also assume detunings to be on the order of 100\;Hz, which is consistent with atom clouds lensed to sub-nK effective temperatures \cite{Deppner2021,Kovachy2015Lensing,abe2021matter} that are desirable for systematic error mitigation \cite{abe2021matter}.  A 0.1 polarization amplitude error gives a reasonable $1\%$ of optical power in unwanted components.  In general, the amplitudes of the $\sigma_+$ and $\sigma_-$ polarization components can be complex.  While Fig. \ref{fig:lineprofile} plots values of these amplitudes over a real domain to ease visualization, we note that using complex values for these coefficients does not significantly affect the scale of the infidelities. We also evaluate the impact of variations in the bias magnetic field on the infidelity compared with other noise channels.

\begin{figure}
    \centering
    \includegraphics[width=\linewidth]{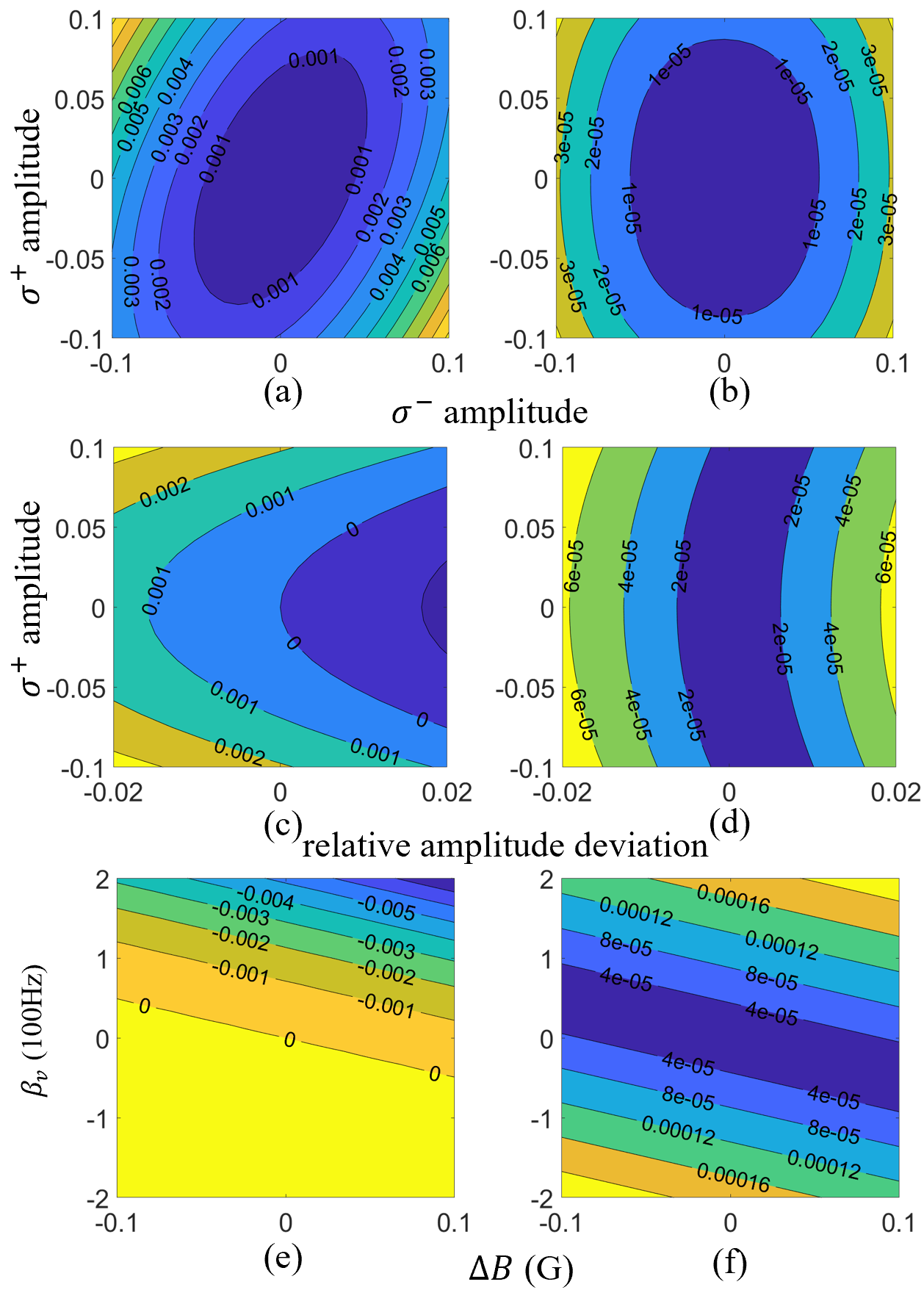}
\caption{Phase deviation induced by single optimized pulse. In the optimized pulse, time-dependent noises are included (the root mean square (RMS) of the noises is $0.5\%$ in the amplitude and $1\%$ in the phase). A set of 10 simulations are performed with random time-dependent noises for different noise parameters. (a), (c), (e) are mean values, and (b), (d), (f) are corresponding standard deviations of the simulation set.}
    \label{fig:single pulse}
\end{figure}

In Fig. \ref{fig:single pulse}, we show the phase deviation induced by a single optimized pulse as a function of different noise channels (i.e., the difference between the actual and ideal final state phase value). An atom interferometer with $\sim$1000 optimized pulses would introduce a total $\sim$ rad scale phase deviation, due to the $\sim$\;mrad scale phase deviation of each pulse for the various noise channels. Atomic dark matter and gravitational wave detectors typically look for time varying signals in a particular frequency band, such as 0.3 - 3 Hz \cite{abe2021matter}.  If, for example, the fractional fluctuation of the polarization errors are at the level of $0.1\%/\sqrt{\text{Hz}}$ in this frequency band, the corresponding interferometer phase noise would be at the level of $\sim\text{mrad}/\sqrt{\text{Hz}}$. Such phase fluctuations may be further suppressed if both arms of the interferometer, or both interferometers in a differential gradiometer configuration \cite{abe2021matter}, experience close to the same errors.

\begin{figure}
    \centering
    \includegraphics[width=\linewidth]{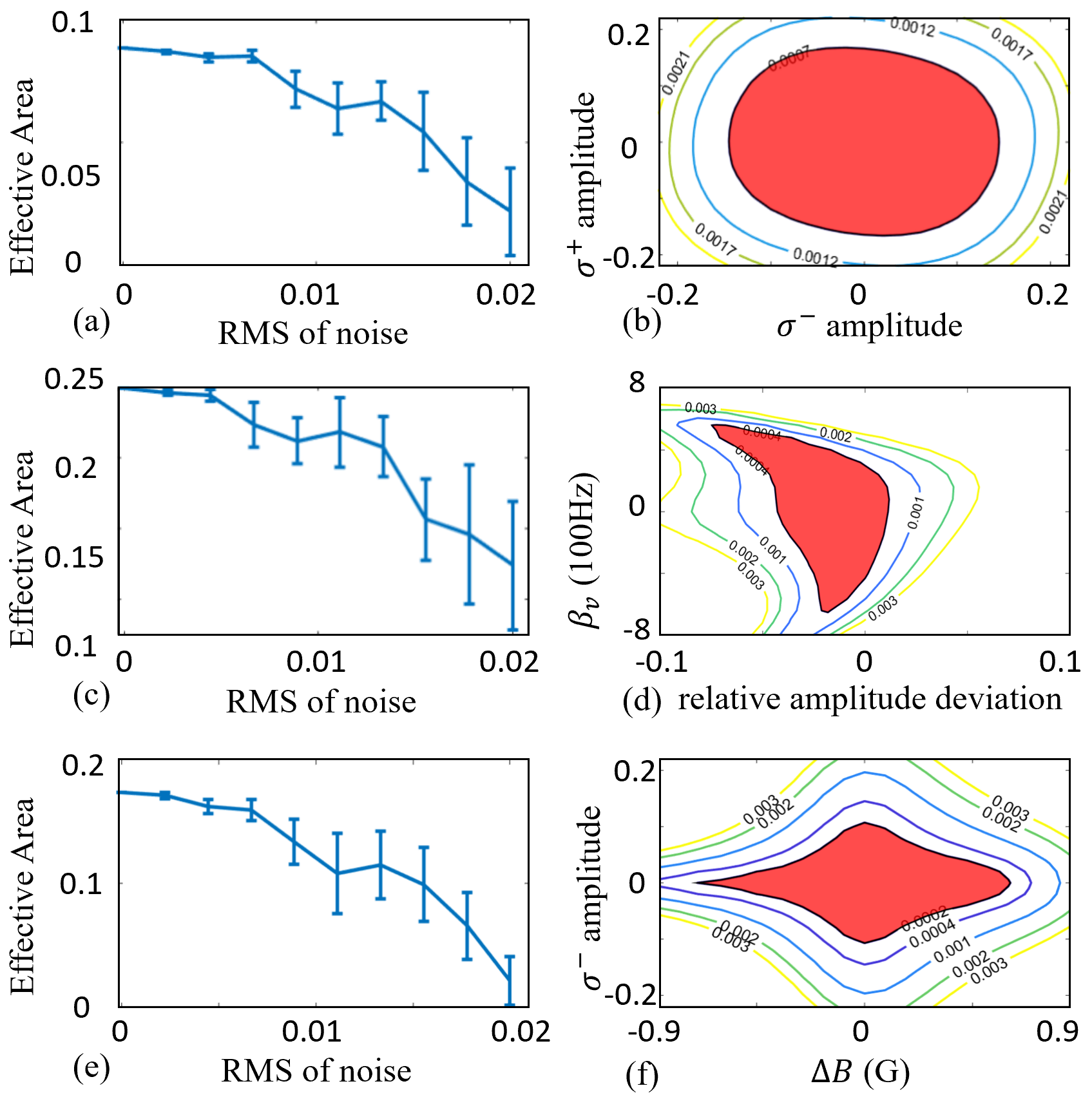}
    \caption{Time-dependent noise impact on infidelities. For the same optimized pulse, adding noise in the control pulse will negatively impact the robustness which is determined by measuring the effective area (red) where the infidelity is a certain value (0.0007 for the polarization map (b), 0.0004 for the amplitude/detuning ($\beta_v$) map (d) and 0.0002 for the $\sigma_-$ polarization/field deviation $\Delta B$ map (f). The robustness on the polarizations [(a)], the amplitude/detuning [(c)], and polarization/magnetic field [(e)] are worse if RMS of the noise increases. A value $x$ of the RMS of noise on the horizontal axis of the plot corresponds to RMS fractional amplitude noise of $x$ and RMS phase noise of $x$ rad. We note that the response to noise is not perfectly symmetric. We do not necessarily expect the optimization to provide perfect symmetry here as we might for a two-level system since the dynamics are significantly more complex, with many different states that have different coupling strengths and different detunings.}
    \label{fig:noise}
\end{figure}

\begin{figure}
    \centering
    \includegraphics[width=\linewidth]{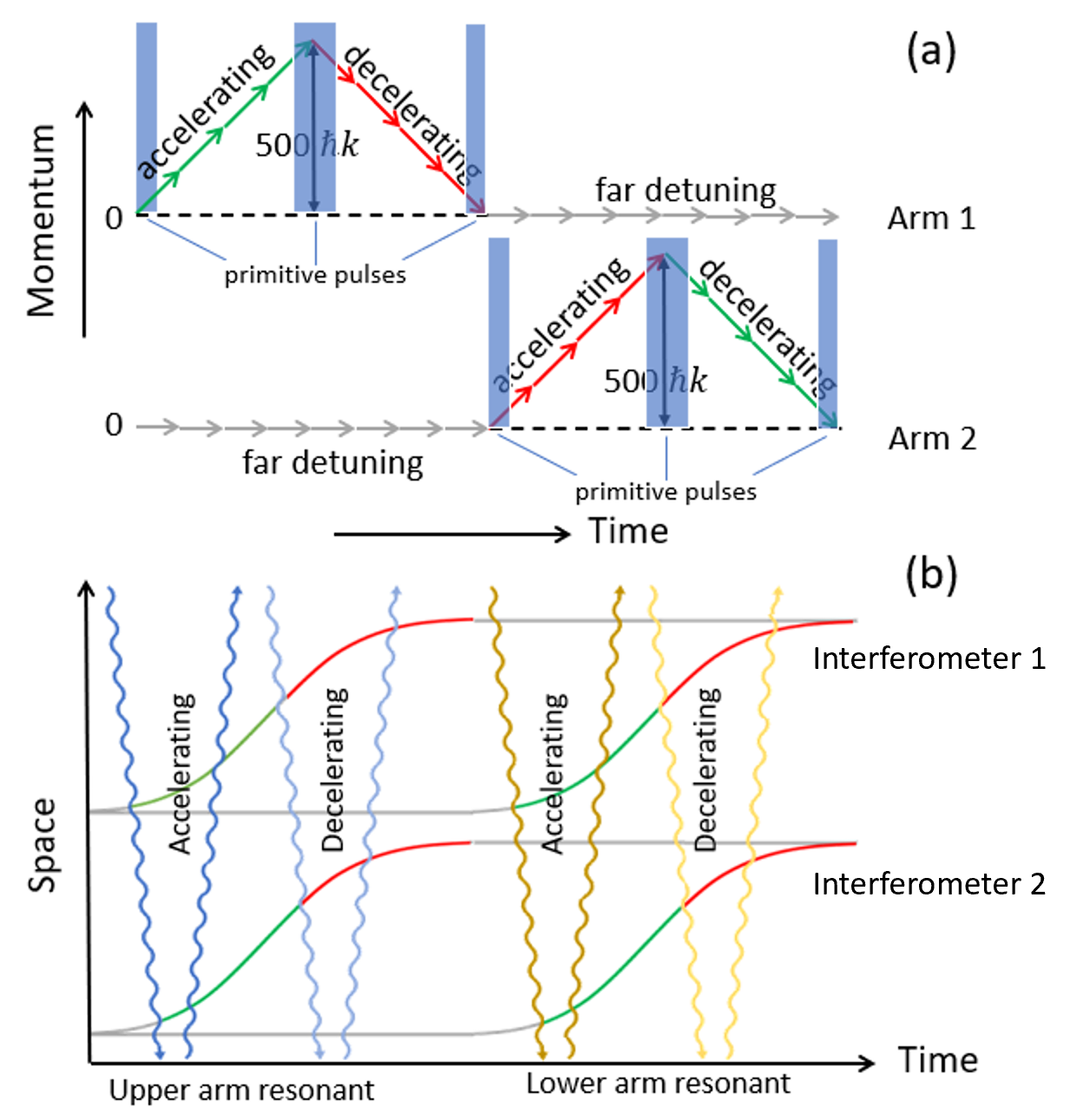}
    \caption{(a) A schematic diagram of momentum change in both arms of the interferometer. Primitive pulses are used in the highlighted parts to avoid unwanted off-resonant interaction with the untransferred arm. The maximum momentum in our simulation setup is 500 $\hbar k$. (b) A schematic diagram of a differential measurement between two interferometers spatially separated over a baseline.}
    \label{fig:schematic}
\end{figure}
 
Since time-dependent noise in the laser amplitude and phase will also occur in the physical system, we studied its impact on the infidelity of the optimized pulse. The added white noise is constructed separately for amplitude and phase by
\begin{equation}
    f(t) =\sum_{i}^{1000} A_i\cos(\omega_i t)+B_i\sin(\omega_i t),
    \label{eqn:time noise}
\end{equation}
where $A_i$ and $B_i$ are randomly generated amplitudes and $\omega_i$ are random frequencies uniformly distributed from 50 Hz to $10^5$ Hz. We normalize each $f(t)$ to the desired root mean square (RMS) value.  The spectrum of the noise is chosen in this range since lower or higher frequencies are tested to have less impact on the robustness. A thousand frequency components are used to guarantee sufficient sampling density in the targeted noise spectrum. Higher sampling numbers do not change the result significantly. Due to the randomly generated noise for each simulation, the impact on the infidelity undergoes fluctuations between different simulation runs. We measure this impact by observing the trend of the area in the contour plot whose infidelity is under a certain value (effective area). By repeating the simulation, the impact fluctuations are indicated by the standard deviation as error bars, and the trend is shown in Fig. \ref{fig:noise}(a), (c), (e). Overall, with bigger noise both in amplitude and phase, the effective area is smaller in the infidelity map for different noise channels (Fig. \ref{fig:noise}(b), (d), (f)).  For the noise RMS of approximately 1 percent (defined in Fig. \ref{fig:noise} caption) which is an achievable level in experiments \footnote{For instance, using commercially available laser systems from Menlo Systems}, the effective area reduction is relatively modest (a factor of approximately 2 or less), indicating that the optimized pulses can remain effective in the presence of noises.  This is further verified by the full interferometer simulation described below, which shows that good total transfer efficiency (including the net effect of all pulses) and well-controlled phase deviations can be maintained for experimentally relevant combinations of static and time-dependent noises, even for interferometers with thousands of total pulses.

\begin{figure*}
    \centering
    \includegraphics[width=\textwidth]{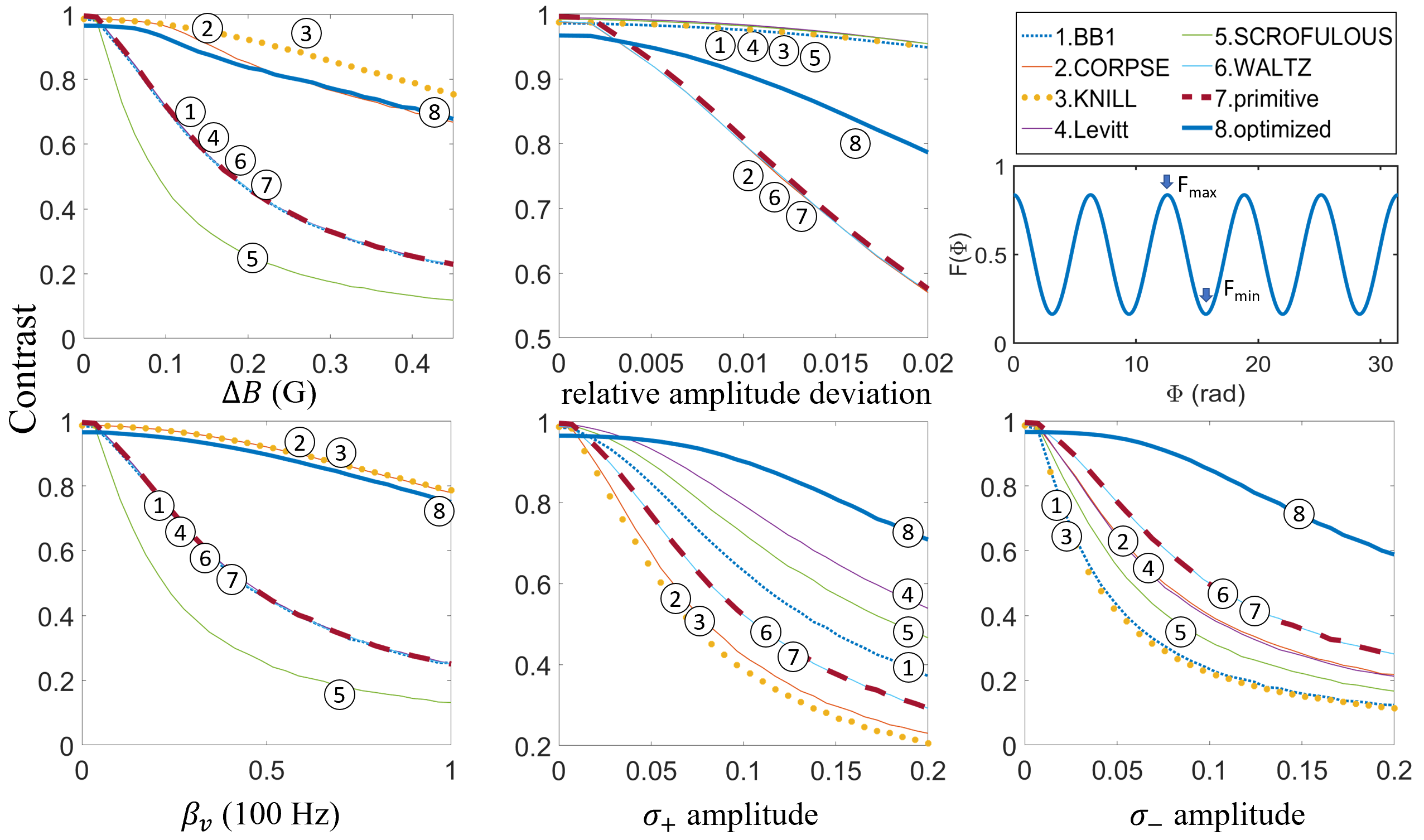}
    \caption{The impact of different noise channels on interference contrast in a full interferometer. The time-dependent noise level is the same as in Figs. 3 and 6. In each panel, the vertical axis is the interference contrast and the horizontal axis is the width $w$ of the zero-centered Gaussian probability distribution from which the corresponding noise parameter $x_{i}$ is sampled. An example of a simulated interference pattern is shown (at the upper-right corner, below the legend) to illustrate the contrast calculation. In contrast to other pulses, the optimized pulse maintains a high degree of robustness against all noise channels, and the optimized pulse outperforms all other pulses in robustness against polarization noise.}
    \label{fig:contrast}
\end{figure*}

\subsection{Full interferometer}
The optimized pulse has been tested in a full atom interferometer simulation (Fig. \ref{fig:schematic}(a)) with total 2000 pulses (each arm experiences 1000 resonant pulses and 1000 off-resonant pulses) as a final feasibility test. In the simulation, we calculate the finite time step unitary evolution of the states using
\begin{eqnarray}
    &u^{(j)} = e^{-i\hat{H}(t_{j})\Delta t_{j}/\hbar}, \\
    &\Psi_{f} =u^{(M)}...u^{(2)}u^{(1)}\Psi_{i},
    \label{chain}
\end{eqnarray}
where $\Psi_{i}$ and $\Psi_{f}$ are initial and final states. In the finite time approximation, the Hamiltonian $\hat{H}(t_j)$ keeps the same from $t_j$ to $t_j+\Delta t_j$, so that the evolution operator for the finite time $\Delta t_j$ is $u^{(j)}$. We calculate the full evolution by applying a chain of finite time operators (Eq. (\ref{chain})). The optimized pulse is divided into $M =2500$ segments (step time is 0.8 $\mu$s, further time step decrease does not significantly influence the results). Since the switching on and off are faster than pulse variations, 50 segments are specially assigned on each switching process to assure their smoothness (see Fig. \ref{fig:setup} (c)). 
In our LMT scheme for clock interferometry \cite{graham2013new} (Fig. \ref{fig:schematic}(a)), we utilize 20 primitive pulses for initial momentum splitting, 480 optimized pulses for continuing the atoms' acceleration, 480 optimized pulses for deceleration, and 20 primitive pulses for symmetrical deceleration. Then, the laser is tuned to be resonant with the other arm, and the same process is repeated on the other arm. During the LMT process, pulses are applied from alternating directions \cite{graham2013new}. The arms are recombined at the end, and we measure the total transfer efficiency and the phase difference between the two arms. 
The reason for including the primitive pulses is that, in creating the full interferometer, the optimized pulses have unwanted off-resonant interaction with the untransferred arm when the relative Doppler shift between the two arms is small. This effect can be reduced by using primitive pulses with appropriate Rabi frequency at small momentum separation. As the relative velocity between the arms increases, the increasing relative Doppler shift causes the pulses to become further detuned from the other arm, and the deleterious effect on the untransferred arm decreases. At this point, we switch to the optimized pulse to boost the momentum splitting to large values.

The time-dependent noise composed from Eq.~(\ref{eqn:time noise}) varies in each pulse in one simulation run. We keep it the same for different simulation runs within a single static noise channel scan, in which we vary two static noise parameters while keeping others the same (Fig.~\ref{fig:interferometer}). 
We carry out repeated static noise channel scans, each with randomly generated time-dependent noise, to evaluate the phase difference between the interferometer arms.  Typical applications of clock interferometry involve differential measurements between two atom interferometers in a gradiometer configuration using a common laser (see Fig. \ref{fig:schematic}(b)) \cite{graham2013new, Arvanitaki2018_darkmatter}. Therefore, in each scan, a differential phase is calculated by subtracting out the phase of an interferometer with zero static noise (i.e. set the center of the map as zero). The mean value and the standard deviation of the differential phase are computed for each static noise point.  If two interferometers in a gradiometer experience the same static noise, the influence of the time-dependent noise on the phase will completely cancel as a common mode \cite{graham2013new}.  By comparing different static noise points, one can determine the residual differential phase noise if the two interferometers in the differential measurement are not matched perfectly in terms of polarization, amplitude, detuning, and magnetic field.

To demonstrate the advantage of the optimized pulse, following a similar approach to other papers studying the application of quantum optimal control to atom interferometry in different contexts \cite{saywell2018_QuOC_Mirror, Goerz2023}, the contrast of the interference pattern after the full interferometer (Fig. \ref{fig:contrast}) is compared with primitive and other composite pulses. The interference pattern is a probability summation of the single-atom interference patterns from an ensemble of atoms. Each atom in the ensemble experienced different static noise according to a particular distribution. For a certain noise channel, we use a Monte Carlo method to construct the interference pattern $F(\Phi)$ by
\begin{eqnarray}
   &f_{i}(\Phi)=   \frac{1}{2}+\frac{1}{2}T(x_{i})\cos\left( \Phi + \Delta \phi(x_{i}) \right),\\
   &F(\Phi)= \frac{1}{N}\sum^{N}_{i=1} f_{i}(\Phi),
\end{eqnarray}
where $\Phi$ is in a range bigger than $2\pi$ to show the interference pattern contrast. In a single-atom interference pattern $f_{i}(\Phi)$, we assign a Gaussian probability distribution whose width is $w$ and pick noise parameter $x_{i}$ randomly according to the distribution. The transfer efficiency $T(x_{i})$ is the probability of atoms being in the correct state to contribute to the contrast (for example, Fig. \ref{fig:interferometer} (a), (d), (g)). The phase deviation $\Delta \phi(x_{i})$ is a value sampled from a Gaussian distribution whose mean value and standard deviation are determined by the noise parameter $x_{i}$ (for example, Fig. \ref{fig:interferometer} (b), (e), (h) for mean value, and (c), (f), (i) for standard deviation). The sampling number $N=50000$ is big enough to give small statistical fluctuation in determining the contrast. The normalized interference pattern $F(\Phi)$ (an example is shown in Fig. \ref{fig:contrast}) is the probability summation of a large number of $f_{i}(\Phi)$. The contrast we measure is
\begin{equation}
    C = F_{max} -  F_{min}.
\end{equation}
where $F_{max}$ and $F_{min}$ are the maximum and the minimum values of $F(\Phi)$.
The contrast value $C$ and its change with the width $w$ for every static noise channel are presented in Fig. \ref{fig:contrast}.

\section{Conclusion}
In summary, we composed an optimized pulse for ${}^{87}\rm{Sr}$ clock interferometry to improve the robustness against multiple noise channels using the QOC method. This work demonstrates the promise of quantum optimal control for extending the scientific reach of strontium clock atom interferometers, potentially paving the way for these interferometers to detect gravitational waves at currently unexplored frequencies and wavelike dark matter. We will study the experimental implementation of the optimized pulses in the future. To achieve the best performance, it may prove valuable to tailor the noise model used in the optimization algorithm to specific, experimentally measured noise properties and spectra. The application of closed-loop quantum optimal control \cite{Feng2018}, in which experimental measurements of pulse fidelities guide the optimization process, to atom interferometry has the potential to offer further improvements.

\begin{acknowledgements}
We thank Jens Koch, Yunwei Lu, and Q-CTRL for valuable discussions.  This material is based upon work supported by the U.S. Department of Energy, Office of Science, National Quantum Information Science Research Centers, Superconducting Quantum Materials and Systems Center (SQMS) under contract number DE-AC02-07CH11359. 
\end{acknowledgements}

\newpage
\appendix
% \onecolumngrid
\section{System Hamiltonian}
We break the Hamiltonian in a rotating frame into two constant terms $H_B$ (Zeeman shifts) and $H_D$ (Doppler shifts) corresponding to free evolution, and a time-dependent control term $H_c$:
\begin{align}
    H(\boldsymbol\beta) = H_B(\beta_B) + H_D(\beta_v) + H_c(\epsilon_+, \epsilon_-, \beta_A),
\end{align}
where $\beta_v$ changes the Doppler shift, $\beta_B$ is the fractional change of the magnetic field, $\beta_A$ is a complex coefficient that characterizes the deviation of the amplitude and the phase of the actual control drive from those of the ideal control drive, and $\epsilon_{\pm}$ are fractional amplitudes of $\sigma_{\pm}$ polarization terms. 

 The Hamiltonian term denoted as $H_B$ can be mathematically divided into two distinct components. The first component is responsible for determining the energy splitting of the ${}^3P_0$ and ${}^1S_0$ hyperfine levels, assuming that the laser is resonant with the transition from $|{}^1S_0;9/2\rangle$ to $|{}^3P_0;9/2\rangle$. This resonance condition ensures that the energies of these two levels are equal in the rotating frame and are therefore set to zero in the Hamiltonian. The energy ladder of the sublevels can be expressed as follows:
\begin{equation}
\left\langle m_F; {}^1{\rm S}_0\left|H_{B1}\right|{}^1{\rm S}_0; m_F\right\rangle
-B \hbar\mu_0 g_S \left(\frac{9}{2}-m_F\right),
\end{equation}

\begin{equation}
\left\langle m_F; {}^3{\rm P}_0\left|H_{B1}\right|{}^3{\rm P}_0; m_F\right\rangle
=
-B \hbar\mu_0 g_P \left(\frac{9}{2}-m_F\right),
\end{equation}
where $B=B_0+ \Delta B = (1+\beta_B)B_0$ is the total magnetic field, $m_F$ is the magnetic quantum number, $\hbar$ is the reduced Planck constant, $\mu_0$ is the vacuum permeability, and $g_S$, $g_P$ are the Landé g-factors for the ${}^1S_0$ and ${}^3P_0$ states, respectively. 
To fully evaluate robustness against magnetic field changes $\Delta B$, we consider the impact of such a magnetic field change for a fixed laser frequency.  As a result of the associated Zeeman shift, the magnetic field change causes the laser to become slightly off-resonant by inducing additional energy shifts of $\Delta B g_S\mu_0 9/2$ and $\Delta B g_P\mu_0 9/2$ for the $|{}^1S_0;9/2\rangle$ and $|{}^3P_0;9/2\rangle$ energy levels, respectively. Consequently, it becomes necessary to introduce a second component in the Hamiltonian to account for these energy shifts. To maintain clarity, the energy shift is symmetrically divided between the ground and excited states, as only the total energy difference $\Delta E = \Delta B g_P\mu_0 9/2 -\Delta B g_S\mu_0 9/2 \equiv \beta_B B_0 \hbar \omega_B$ between the two levels affects the evolution results. Therefore, the complete Hamiltonian term $H_B$ can be written as

\begin{equation}
\begin{split}
\left\langle m_F; {}^1{\rm S}_0\left|H_B\right|{}^1{\rm S}_0; m_F\right\rangle
=-\beta_B B_0 \frac{ \hbar  \omega_B}{2} \\
-\left(1+\beta_B\right) B_0 \hbar\mu_0 g_S \left(\frac{9}{2}-m_F\right),
\end{split}
\end{equation}

\begin{equation}
\begin{split}
\left\langle m_F; {}^3 {\rm P}_0\left|H_B\right|{}^3{\rm P}_0 ;m_F\right\rangle
=\beta_B B_0\frac{ \hbar \omega_B}{2}\\
-\left(1+\beta_B\right) B_0 \hbar\mu_0 g_P \left(\frac{9}{2}-m_F\right),
\end{split}
\end{equation}
where $B_0=1$ G is the nominal bias field and $\omega_B = 2\pi \times 491 $ Hz/G. In the second term $\mu_0 g_S = 2\pi \times 182 $ Hz/G, and $\mu_0 g_P = 2\pi \times 291$ Hz/G \cite{boydthesis}. This term gives the energy ladder of all sublevels with a fractional field change of $\beta_B$.  

Similarly, the $H_D$ matrix elements are given by

\begin{equation}
\left\langle m_F; {}^1{\rm S}_0\left|\hat{H}_D\right|{}^1{\rm S}_0 ;m_F\right\rangle=-\beta_v\hbar \omega_D / 2,
\end{equation}

\begin{equation}
\left\langle m_F; {}^3{\rm P}_0\left|\hat{H}_D\right|{}^3{\rm P}_0 ;m_F\right\rangle=\beta_v\hbar \omega_D / 2,
\end{equation}
where $\omega_D=2\pi\times 100$ $\rm{Hz}$ is chosen as a factorization so that $\beta_v$ is in units of 100 Hz in all the figures. This term accounts for the energy shift due to the Doppler effect. The sign of $H_D$ term will flip in the rotating frame when the laser pulse direction switches. 

For the laser control Hamiltonian term $H_C$, the matrix elements are 

\begin{equation}
\left\langle m_F; {}^3 {\rm P}_0\left|\hat{H}_C\right|{}^1 {\rm S}_0 ;m_F\right\rangle=\left(1+\beta_A \right) \frac{C^\pi_{m_F}}{C^\pi_{9/2}} \frac{\hbar\Omega(t) \sqrt{1-{\epsilon}^2}}{2},
\end{equation}

\begin{equation}
\left\langle m_F+1; {}^3 {\rm P}_0\left|\hat{H}_C\right|{}^1 {\rm S}_0 ;m_F\right\rangle=\left(1+\beta_A \right)\frac{C^+_{m_F}}{C^\pi_{9/2}} \frac{\hbar\Omega(t) \epsilon_{+}}{2} ,
\end{equation}

\begin{equation}
\left\langle m_F-1; {}^3 {\rm P}_0\left|\hat{H}_C\right|{}^1 {\rm S}_0 ;m_F\right\rangle=\left(1+\beta_A \right)\frac{C^-_{m_F}}{C^\pi_{9/2}}  \frac{\hbar\Omega(t)\epsilon_{-}}{2} ,
\end{equation}
where ratios $C^\pi_{m_F}$, $C^+_{m_F}$, $C^-_{m_F}$ are Clebsch-Gordon coefficients accounting for the different transition strengths which are derived from the electric dipole matrix element between the $^1{\rm S}_0$ and $^3{\rm P}_0$ states. These elements arise from an admixture between $^1{\rm P}_1$ and $^3{\rm P}_0$ states due to spin-orbit coupling and hyperfine interactions (the dipole matrix elements  $\left\langle {}^3 {\rm P}_0\left|\hat{H}_C\right|{}^1 {\rm S}_0 \right\rangle$ are proportional to $\left\langle {}^1 {\rm P}_1\left|\hat{H}_C\right|{}^1 {\rm S}_0 \right\rangle$) \cite{boydthesis}. The calculation of such electric dipole moments is described, for example, in Metcalf and Van der Straten's book \cite{metcalf1999laser}. $\beta_A(t)$ is composed of a constant deviation $\beta_{A0}$ and white noises on both amplitude and phase (see Eq. (\ref{eqn:time noise})). $\Omega(t)$ is the Rabi frequency which varies with the control pulse amplitude, and its peak value is $2\pi\times 3 \times 10^ 3$ Hz. $\epsilon^2 = |\epsilon_-|^2+|\epsilon_+|^2$ conserves the total amplitude. All other matrix elements not explicitly listed are zero.

% \bibliographystyle{apsrev4-2}
% \bibliography{QOCSubmit.bib}

\begin{thebibliography}{94}%
\makeatletter
\providecommand \@ifxundefined [1]{%
 \@ifx{#1\undefined}
}%
\providecommand \@ifnum [1]{%
 \ifnum #1\expandafter \@firstoftwo
 \else \expandafter \@secondoftwo
 \fi
}%
\providecommand \@ifx [1]{%
 \ifx #1\expandafter \@firstoftwo
 \else \expandafter \@secondoftwo
 \fi
}%
\providecommand \natexlab [1]{#1}%
\providecommand \enquote  [1]{``#1''}%
\providecommand \bibnamefont  [1]{#1}%
\providecommand \bibfnamefont [1]{#1}%
\providecommand \citenamefont [1]{#1}%
\providecommand \href@noop [0]{\@secondoftwo}%
\providecommand \href [0]{\begingroup \@sanitize@url \@href}%
\providecommand \@href[1]{\@@startlink{#1}\@@href}%
\providecommand \@@href[1]{\endgroup#1\@@endlink}%
\providecommand \@sanitize@url [0]{\catcode `\\12\catcode `\$12\catcode
  `\&12\catcode `\#12\catcode `\^12\catcode `\_12\catcode `\%12\relax}%
\providecommand \@@startlink[1]{}%
\providecommand \@@endlink[0]{}%
\providecommand \url  [0]{\begingroup\@sanitize@url \@url }%
\providecommand \@url [1]{\endgroup\@href {#1}{\urlprefix }}%
\providecommand \urlprefix  [0]{URL }%
\providecommand \Eprint [0]{\href }%
\providecommand \doibase [0]{https://doi.org/}%
\providecommand \selectlanguage [0]{\@gobble}%
\providecommand \bibinfo  [0]{\@secondoftwo}%
\providecommand \bibfield  [0]{\@secondoftwo}%
\providecommand \translation [1]{[#1]}%
\providecommand \BibitemOpen [0]{}%
\providecommand \bibitemStop [0]{}%
\providecommand \bibitemNoStop [0]{.\EOS\space}%
\providecommand \EOS [0]{\spacefactor3000\relax}%
\providecommand \BibitemShut  [1]{\csname bibitem#1\endcsname}%
\let\auto@bib@innerbib\@empty
%</preamble>
\bibitem [{\citenamefont {Zhou}\ \emph {et~al.}(2015)\citenamefont {Zhou},
  \citenamefont {Long}, \citenamefont {Tang}, \citenamefont {Chen},
  \citenamefont {Gao}, \citenamefont {Peng}, \citenamefont {Duan},
  \citenamefont {Zhong}, \citenamefont {Xiong}, \citenamefont {Wang},
  \citenamefont {Zhang},\ and\ \citenamefont {Zhan}}]{Zhou2015_equivalence}%
  \BibitemOpen
  \bibfield  {author} {\bibinfo {author} {\bibfnamefont {L.}~\bibnamefont
  {Zhou}}, \bibinfo {author} {\bibfnamefont {S.}~\bibnamefont {Long}}, \bibinfo
  {author} {\bibfnamefont {B.}~\bibnamefont {Tang}}, \bibinfo {author}
  {\bibfnamefont {X.}~\bibnamefont {Chen}}, \bibinfo {author} {\bibfnamefont
  {F.}~\bibnamefont {Gao}}, \bibinfo {author} {\bibfnamefont {W.}~\bibnamefont
  {Peng}}, \bibinfo {author} {\bibfnamefont {W.}~\bibnamefont {Duan}}, \bibinfo
  {author} {\bibfnamefont {J.}~\bibnamefont {Zhong}}, \bibinfo {author}
  {\bibfnamefont {Z.}~\bibnamefont {Xiong}}, \bibinfo {author} {\bibfnamefont
  {J.}~\bibnamefont {Wang}}, \bibinfo {author} {\bibfnamefont {Y.}~\bibnamefont
  {Zhang}},\ and\ \bibinfo {author} {\bibfnamefont {M.}~\bibnamefont {Zhan}},\
  }\href {https://doi.org/10.1103/PhysRevLett.115.013004} {\bibfield  {journal}
  {\bibinfo  {journal} {Phys. Rev. Lett.}\ }\textbf {\bibinfo {volume} {115}},\
  \bibinfo {pages} {013004} (\bibinfo {year} {2015})}\BibitemShut {NoStop}%
\bibitem [{\citenamefont {Rosi}\ \emph {et~al.}(2017)\citenamefont {Rosi},
  \citenamefont {D’Amico}, \citenamefont {Cacciapuoti}, \citenamefont
  {Sorrentino}, \citenamefont {Prevedelli}, \citenamefont {Zych}, \citenamefont
  {Brukner},\ and\ \citenamefont {Tino}}]{rosi2017equivalence}%
  \BibitemOpen
  \bibfield  {author} {\bibinfo {author} {\bibfnamefont {G.}~\bibnamefont
  {Rosi}}, \bibinfo {author} {\bibfnamefont {G.}~\bibnamefont {D’Amico}},
  \bibinfo {author} {\bibfnamefont {L.}~\bibnamefont {Cacciapuoti}}, \bibinfo
  {author} {\bibfnamefont {F.}~\bibnamefont {Sorrentino}}, \bibinfo {author}
  {\bibfnamefont {M.}~\bibnamefont {Prevedelli}}, \bibinfo {author}
  {\bibfnamefont {M.}~\bibnamefont {Zych}}, \bibinfo {author} {\bibfnamefont
  {{\v{C}}.}~\bibnamefont {Brukner}},\ and\ \bibinfo {author} {\bibfnamefont
  {G.}~\bibnamefont {Tino}},\ }\href@noop {} {\bibfield  {journal} {\bibinfo
  {journal} {Nat. Comm.}\ }\textbf {\bibinfo {volume} {8}},\ \bibinfo {pages}
  {1} (\bibinfo {year} {2017})}\BibitemShut {NoStop}%
\bibitem [{\citenamefont {Overstreet}\ \emph {et~al.}(2018)\citenamefont
  {Overstreet}, \citenamefont {Asenbaum}, \citenamefont {Kovachy},
  \citenamefont {Notermans}, \citenamefont {Hogan},\ and\ \citenamefont
  {Kasevich}}]{Overstreet2018_equivalence}%
  \BibitemOpen
  \bibfield  {author} {\bibinfo {author} {\bibfnamefont {C.}~\bibnamefont
  {Overstreet}}, \bibinfo {author} {\bibfnamefont {P.}~\bibnamefont
  {Asenbaum}}, \bibinfo {author} {\bibfnamefont {T.}~\bibnamefont {Kovachy}},
  \bibinfo {author} {\bibfnamefont {R.}~\bibnamefont {Notermans}}, \bibinfo
  {author} {\bibfnamefont {J.~M.}\ \bibnamefont {Hogan}},\ and\ \bibinfo
  {author} {\bibfnamefont {M.~A.}\ \bibnamefont {Kasevich}},\ }\href
  {https://doi.org/10.1103/PhysRevLett.120.183604} {\bibfield  {journal}
  {\bibinfo  {journal} {Phys. Rev. Lett.}\ }\textbf {\bibinfo {volume} {120}},\
  \bibinfo {pages} {183604} (\bibinfo {year} {2018})}\BibitemShut {NoStop}%
\bibitem [{\citenamefont {Asenbaum}\ \emph {et~al.}(2017)\citenamefont
  {Asenbaum}, \citenamefont {Overstreet}, \citenamefont {Kovachy},
  \citenamefont {Brown}, \citenamefont {Hogan},\ and\ \citenamefont
  {Kasevich}}]{Asenbaum2017}%
  \BibitemOpen
  \bibfield  {author} {\bibinfo {author} {\bibfnamefont {P.}~\bibnamefont
  {Asenbaum}}, \bibinfo {author} {\bibfnamefont {C.}~\bibnamefont
  {Overstreet}}, \bibinfo {author} {\bibfnamefont {T.}~\bibnamefont {Kovachy}},
  \bibinfo {author} {\bibfnamefont {D.~D.}\ \bibnamefont {Brown}}, \bibinfo
  {author} {\bibfnamefont {J.~M.}\ \bibnamefont {Hogan}},\ and\ \bibinfo
  {author} {\bibfnamefont {M.~A.}\ \bibnamefont {Kasevich}},\ }\href
  {https://doi.org/10.1103/PhysRevLett.118.183602} {\bibfield  {journal}
  {\bibinfo  {journal} {Phys. Rev. Lett.}\ }\textbf {\bibinfo {volume} {118}},\
  \bibinfo {pages} {183602} (\bibinfo {year} {2017})}\BibitemShut {NoStop}%
\bibitem [{\citenamefont {Asenbaum}\ \emph {et~al.}(2020)\citenamefont
  {Asenbaum}, \citenamefont {Overstreet}, \citenamefont {Kim}, \citenamefont
  {Curti},\ and\ \citenamefont {Kasevich}}]{Asenbaum2020_equivalence}%
  \BibitemOpen
  \bibfield  {author} {\bibinfo {author} {\bibfnamefont {P.}~\bibnamefont
  {Asenbaum}}, \bibinfo {author} {\bibfnamefont {C.}~\bibnamefont
  {Overstreet}}, \bibinfo {author} {\bibfnamefont {M.}~\bibnamefont {Kim}},
  \bibinfo {author} {\bibfnamefont {J.}~\bibnamefont {Curti}},\ and\ \bibinfo
  {author} {\bibfnamefont {M.~A.}\ \bibnamefont {Kasevich}},\ }\href
  {https://doi.org/10.1103/PhysRevLett.125.191101} {\bibfield  {journal}
  {\bibinfo  {journal} {Phys. Rev. Lett.}\ }\textbf {\bibinfo {volume} {125}},\
  \bibinfo {pages} {191101} (\bibinfo {year} {2020})}\BibitemShut {NoStop}%
\bibitem [{\citenamefont {Kovachy}\ \emph
  {et~al.}(2015{\natexlab{a}})\citenamefont {Kovachy}, \citenamefont
  {Asenbaum}, \citenamefont {Overstreet}, \citenamefont {Donnelly},
  \citenamefont {Dickerson}, \citenamefont {Sugarbaker}, \citenamefont
  {Hogan},\ and\ \citenamefont {Kasevich}}]{kovachy2015_halfmeter}%
  \BibitemOpen
  \bibfield  {author} {\bibinfo {author} {\bibfnamefont {T.}~\bibnamefont
  {Kovachy}}, \bibinfo {author} {\bibfnamefont {P.}~\bibnamefont {Asenbaum}},
  \bibinfo {author} {\bibfnamefont {C.}~\bibnamefont {Overstreet}}, \bibinfo
  {author} {\bibfnamefont {C.}~\bibnamefont {Donnelly}}, \bibinfo {author}
  {\bibfnamefont {S.}~\bibnamefont {Dickerson}}, \bibinfo {author}
  {\bibfnamefont {A.}~\bibnamefont {Sugarbaker}}, \bibinfo {author}
  {\bibfnamefont {J.}~\bibnamefont {Hogan}},\ and\ \bibinfo {author}
  {\bibfnamefont {M.}~\bibnamefont {Kasevich}},\ }\href@noop {} {\bibfield
  {journal} {\bibinfo  {journal} {Nature}\ }\textbf {\bibinfo {volume} {528}},\
  \bibinfo {pages} {530} (\bibinfo {year} {2015}{\natexlab{a}})}\BibitemShut
  {NoStop}%
\bibitem [{\citenamefont {Fray}\ \emph {et~al.}(2004)\citenamefont {Fray},
  \citenamefont {Diez}, \citenamefont {H\"ansch},\ and\ \citenamefont
  {Weitz}}]{fray2004atomic}%
  \BibitemOpen
  \bibfield  {author} {\bibinfo {author} {\bibfnamefont {S.}~\bibnamefont
  {Fray}}, \bibinfo {author} {\bibfnamefont {C.~A.}\ \bibnamefont {Diez}},
  \bibinfo {author} {\bibfnamefont {T.~W.}\ \bibnamefont {H\"ansch}},\ and\
  \bibinfo {author} {\bibfnamefont {M.}~\bibnamefont {Weitz}},\ }\href
  {https://doi.org/10.1103/PhysRevLett.93.240404} {\bibfield  {journal}
  {\bibinfo  {journal} {Phys. Rev. Lett.}\ }\textbf {\bibinfo {volume} {93}},\
  \bibinfo {pages} {240404} (\bibinfo {year} {2004})}\BibitemShut {NoStop}%
\bibitem [{\citenamefont {Schlippert}\ \emph {et~al.}(2014)\citenamefont
  {Schlippert}, \citenamefont {Hartwig}, \citenamefont {Albers}, \citenamefont
  {Richardson}, \citenamefont {Schubert}, \citenamefont {Roura}, \citenamefont
  {Schleich}, \citenamefont {Ertmer},\ and\ \citenamefont
  {Rasel}}]{Schlippert2014ep}%
  \BibitemOpen
  \bibfield  {author} {\bibinfo {author} {\bibfnamefont {D.}~\bibnamefont
  {Schlippert}}, \bibinfo {author} {\bibfnamefont {J.}~\bibnamefont {Hartwig}},
  \bibinfo {author} {\bibfnamefont {H.}~\bibnamefont {Albers}}, \bibinfo
  {author} {\bibfnamefont {L.~L.}\ \bibnamefont {Richardson}}, \bibinfo
  {author} {\bibfnamefont {C.}~\bibnamefont {Schubert}}, \bibinfo {author}
  {\bibfnamefont {A.}~\bibnamefont {Roura}}, \bibinfo {author} {\bibfnamefont
  {W.~P.}\ \bibnamefont {Schleich}}, \bibinfo {author} {\bibfnamefont
  {W.}~\bibnamefont {Ertmer}},\ and\ \bibinfo {author} {\bibfnamefont {E.~M.}\
  \bibnamefont {Rasel}},\ }\href
  {https://doi.org/10.1103/PhysRevLett.112.203002} {\bibfield  {journal}
  {\bibinfo  {journal} {Phys. Rev. Lett.}\ }\textbf {\bibinfo {volume} {112}},\
  \bibinfo {pages} {203002} (\bibinfo {year} {2014})}\BibitemShut {NoStop}%
\bibitem [{\citenamefont {Barrett}\ \emph {et~al.}(2016)\citenamefont
  {Barrett}, \citenamefont {Antoni-Micollier}, \citenamefont {Chichet},
  \citenamefont {Battelier}, \citenamefont {L{\'e}v{\`e}que}, \citenamefont
  {Landragin},\ and\ \citenamefont {Bouyer}}]{Barrett2016}%
  \BibitemOpen
  \bibfield  {author} {\bibinfo {author} {\bibfnamefont {B.}~\bibnamefont
  {Barrett}}, \bibinfo {author} {\bibfnamefont {L.}~\bibnamefont
  {Antoni-Micollier}}, \bibinfo {author} {\bibfnamefont {L.}~\bibnamefont
  {Chichet}}, \bibinfo {author} {\bibfnamefont {B.}~\bibnamefont {Battelier}},
  \bibinfo {author} {\bibfnamefont {T.}~\bibnamefont {L{\'e}v{\`e}que}},
  \bibinfo {author} {\bibfnamefont {A.}~\bibnamefont {Landragin}},\ and\
  \bibinfo {author} {\bibfnamefont {P.}~\bibnamefont {Bouyer}},\ }\href
  {https://doi.org/10.1038/ncomms13786} {\bibfield  {journal} {\bibinfo
  {journal} {Nat. Comm.}\ }\textbf {\bibinfo {volume} {7}},\ \bibinfo {pages}
  {13786} (\bibinfo {year} {2016})}\BibitemShut {NoStop}%
\bibitem [{\citenamefont {Kuhn}\ \emph {et~al.}(2014)\citenamefont {Kuhn},
  \citenamefont {McDonald}, \citenamefont {Hardman}, \citenamefont {Bennetts},
  \citenamefont {Everitt}, \citenamefont {Altin}, \citenamefont {Debs},
  \citenamefont {Close},\ and\ \citenamefont {Robins}}]{kuhn2014bose}%
  \BibitemOpen
  \bibfield  {author} {\bibinfo {author} {\bibfnamefont {C.~C.~N.}\
  \bibnamefont {Kuhn}}, \bibinfo {author} {\bibfnamefont {G.~D.}\ \bibnamefont
  {McDonald}}, \bibinfo {author} {\bibfnamefont {K.~S.}\ \bibnamefont
  {Hardman}}, \bibinfo {author} {\bibfnamefont {S.}~\bibnamefont {Bennetts}},
  \bibinfo {author} {\bibfnamefont {P.~J.}\ \bibnamefont {Everitt}}, \bibinfo
  {author} {\bibfnamefont {P.~A.}\ \bibnamefont {Altin}}, \bibinfo {author}
  {\bibfnamefont {J.~E.}\ \bibnamefont {Debs}}, \bibinfo {author}
  {\bibfnamefont {J.~D.}\ \bibnamefont {Close}},\ and\ \bibinfo {author}
  {\bibfnamefont {N.~P.}\ \bibnamefont {Robins}},\ }\href
  {https://doi.org/10.1088/1367-2630/16/7/073035} {\bibfield  {journal}
  {\bibinfo  {journal} {New J. Phys.}\ }\textbf {\bibinfo {volume} {16}},\
  \bibinfo {pages} {073035} (\bibinfo {year} {2014})}\BibitemShut {NoStop}%
\bibitem [{\citenamefont {Barrett}\ \emph {et~al.}(2015)\citenamefont
  {Barrett}, \citenamefont {Antoni-Micollier}, \citenamefont {Chichet},
  \citenamefont {Battelier}, \citenamefont {Gominet}, \citenamefont {Bertoldi},
  \citenamefont {Bouyer},\ and\ \citenamefont
  {Landragin}}]{barrett2015correlative}%
  \BibitemOpen
  \bibfield  {author} {\bibinfo {author} {\bibfnamefont {B.}~\bibnamefont
  {Barrett}}, \bibinfo {author} {\bibfnamefont {L.}~\bibnamefont
  {Antoni-Micollier}}, \bibinfo {author} {\bibfnamefont {L.}~\bibnamefont
  {Chichet}}, \bibinfo {author} {\bibfnamefont {B.}~\bibnamefont {Battelier}},
  \bibinfo {author} {\bibfnamefont {P.-A.}\ \bibnamefont {Gominet}}, \bibinfo
  {author} {\bibfnamefont {A.}~\bibnamefont {Bertoldi}}, \bibinfo {author}
  {\bibfnamefont {P.}~\bibnamefont {Bouyer}},\ and\ \bibinfo {author}
  {\bibfnamefont {A.}~\bibnamefont {Landragin}},\ }\href
  {https://doi.org/10.1088/1367-2630/17/8/085010} {\bibfield  {journal}
  {\bibinfo  {journal} {New J. Phys.}\ }\textbf {\bibinfo {volume} {17}},\
  \bibinfo {pages} {085010} (\bibinfo {year} {2015})}\BibitemShut {NoStop}%
\bibitem [{\citenamefont {Tarallo}\ \emph {et~al.}(2014)\citenamefont
  {Tarallo}, \citenamefont {Mazzoni}, \citenamefont {Poli}, \citenamefont
  {Sutyrin}, \citenamefont {Zhang},\ and\ \citenamefont
  {Tino}}]{PhysRevLett.113.023005}%
  \BibitemOpen
  \bibfield  {author} {\bibinfo {author} {\bibfnamefont {M.~G.}\ \bibnamefont
  {Tarallo}}, \bibinfo {author} {\bibfnamefont {T.}~\bibnamefont {Mazzoni}},
  \bibinfo {author} {\bibfnamefont {N.}~\bibnamefont {Poli}}, \bibinfo {author}
  {\bibfnamefont {D.~V.}\ \bibnamefont {Sutyrin}}, \bibinfo {author}
  {\bibfnamefont {X.}~\bibnamefont {Zhang}},\ and\ \bibinfo {author}
  {\bibfnamefont {G.~M.}\ \bibnamefont {Tino}},\ }\href
  {https://doi.org/10.1103/PhysRevLett.113.023005} {\bibfield  {journal}
  {\bibinfo  {journal} {Phys. Rev. Lett.}\ }\textbf {\bibinfo {volume} {113}},\
  \bibinfo {pages} {023005} (\bibinfo {year} {2014})}\BibitemShut {NoStop}%
\bibitem [{\citenamefont {Bonnin}\ \emph {et~al.}(2013)\citenamefont {Bonnin},
  \citenamefont {Zahzam}, \citenamefont {Bidel},\ and\ \citenamefont
  {Bresson}}]{PhysRevA.88.043615}%
  \BibitemOpen
  \bibfield  {author} {\bibinfo {author} {\bibfnamefont {A.}~\bibnamefont
  {Bonnin}}, \bibinfo {author} {\bibfnamefont {N.}~\bibnamefont {Zahzam}},
  \bibinfo {author} {\bibfnamefont {Y.}~\bibnamefont {Bidel}},\ and\ \bibinfo
  {author} {\bibfnamefont {A.}~\bibnamefont {Bresson}},\ }\href
  {https://doi.org/10.1103/PhysRevA.88.043615} {\bibfield  {journal} {\bibinfo
  {journal} {Phys. Rev. A}\ }\textbf {\bibinfo {volume} {88}},\ \bibinfo
  {pages} {043615} (\bibinfo {year} {2013})}\BibitemShut {NoStop}%
\bibitem [{\citenamefont {Hartwig}\ \emph {et~al.}(2015)\citenamefont
  {Hartwig}, \citenamefont {Abend}, \citenamefont {Schubert}, \citenamefont
  {Schlippert}, \citenamefont {Ahlers}, \citenamefont {Posso-Trujillo},
  \citenamefont {Gaaloul}, \citenamefont {Ertmer},\ and\ \citenamefont
  {Rasel}}]{Hartwig2015}%
  \BibitemOpen
  \bibfield  {author} {\bibinfo {author} {\bibfnamefont {J.}~\bibnamefont
  {Hartwig}}, \bibinfo {author} {\bibfnamefont {S.}~\bibnamefont {Abend}},
  \bibinfo {author} {\bibfnamefont {C.}~\bibnamefont {Schubert}}, \bibinfo
  {author} {\bibfnamefont {D.}~\bibnamefont {Schlippert}}, \bibinfo {author}
  {\bibfnamefont {H.}~\bibnamefont {Ahlers}}, \bibinfo {author} {\bibfnamefont
  {K.}~\bibnamefont {Posso-Trujillo}}, \bibinfo {author} {\bibfnamefont
  {N.}~\bibnamefont {Gaaloul}}, \bibinfo {author} {\bibfnamefont
  {W.}~\bibnamefont {Ertmer}},\ and\ \bibinfo {author} {\bibfnamefont {E.~M.}\
  \bibnamefont {Rasel}},\ }\href
  {https://doi.org/10.1088/1367-2630/17/3/035011} {\bibfield  {journal}
  {\bibinfo  {journal} {New J. Phys.}\ }\textbf {\bibinfo {volume} {17}},\
  \bibinfo {pages} {035011} (\bibinfo {year} {2015})}\BibitemShut {NoStop}%
\bibitem [{\citenamefont {Williams}\ \emph {et~al.}(2016)\citenamefont
  {Williams}, \citenamefont {wey Chiow}, \citenamefont {Yu},\ and\
  \citenamefont {M\"{u}ller}}]{williams2016quantum}%
  \BibitemOpen
  \bibfield  {author} {\bibinfo {author} {\bibfnamefont {J.}~\bibnamefont
  {Williams}}, \bibinfo {author} {\bibfnamefont {S.}~\bibnamefont {wey Chiow}},
  \bibinfo {author} {\bibfnamefont {N.}~\bibnamefont {Yu}},\ and\ \bibinfo
  {author} {\bibfnamefont {H.}~\bibnamefont {M\"{u}ller}},\ }\href
  {https://doi.org/10.1088/1367-2630/18/2/025018} {\bibfield  {journal}
  {\bibinfo  {journal} {New J. Phys.}\ }\textbf {\bibinfo {volume} {18}},\
  \bibinfo {pages} {025018} (\bibinfo {year} {2016})}\BibitemShut {NoStop}%
\bibitem [{\citenamefont {Overstreet}\ \emph {et~al.}(2022)\citenamefont
  {Overstreet}, \citenamefont {Asenbaum}, \citenamefont {Curti}, \citenamefont
  {Kim},\ and\ \citenamefont {Kasevich}}]{overstreet2022observation}%
  \BibitemOpen
  \bibfield  {author} {\bibinfo {author} {\bibfnamefont {C.}~\bibnamefont
  {Overstreet}}, \bibinfo {author} {\bibfnamefont {P.}~\bibnamefont
  {Asenbaum}}, \bibinfo {author} {\bibfnamefont {J.}~\bibnamefont {Curti}},
  \bibinfo {author} {\bibfnamefont {M.}~\bibnamefont {Kim}},\ and\ \bibinfo
  {author} {\bibfnamefont {M.~A.}\ \bibnamefont {Kasevich}},\ }\href@noop {}
  {\bibfield  {journal} {\bibinfo  {journal} {Science}\ }\textbf {\bibinfo
  {volume} {375}},\ \bibinfo {pages} {226} (\bibinfo {year}
  {2022})}\BibitemShut {NoStop}%
\bibitem [{\citenamefont {Dimopoulos}\ \emph {et~al.}(2008)\citenamefont
  {Dimopoulos}, \citenamefont {Graham}, \citenamefont {Hogan}, \citenamefont
  {Kasevich},\ and\ \citenamefont {Rajendran}}]{Dimopoulos2008_GW}%
  \BibitemOpen
  \bibfield  {author} {\bibinfo {author} {\bibfnamefont {S.}~\bibnamefont
  {Dimopoulos}}, \bibinfo {author} {\bibfnamefont {P.~W.}\ \bibnamefont
  {Graham}}, \bibinfo {author} {\bibfnamefont {J.~M.}\ \bibnamefont {Hogan}},
  \bibinfo {author} {\bibfnamefont {M.~A.}\ \bibnamefont {Kasevich}},\ and\
  \bibinfo {author} {\bibfnamefont {S.}~\bibnamefont {Rajendran}},\ }\href
  {https://doi.org/10.1103/PhysRevD.78.122002} {\bibfield  {journal} {\bibinfo
  {journal} {Phys. Rev. D}\ }\textbf {\bibinfo {volume} {78}},\ \bibinfo
  {pages} {122002} (\bibinfo {year} {2008})}\BibitemShut {NoStop}%
\bibitem [{\citenamefont {Graham}\ \emph {et~al.}(2013)\citenamefont {Graham},
  \citenamefont {Hogan}, \citenamefont {Kasevich},\ and\ \citenamefont
  {Rajendran}}]{graham2013new}%
  \BibitemOpen
  \bibfield  {author} {\bibinfo {author} {\bibfnamefont {P.~W.}\ \bibnamefont
  {Graham}}, \bibinfo {author} {\bibfnamefont {J.~M.}\ \bibnamefont {Hogan}},
  \bibinfo {author} {\bibfnamefont {M.~A.}\ \bibnamefont {Kasevich}},\ and\
  \bibinfo {author} {\bibfnamefont {S.}~\bibnamefont {Rajendran}},\ }\href
  {https://doi.org/10.1103/PhysRevLett.110.171102} {\bibfield  {journal}
  {\bibinfo  {journal} {Phys. Rev. Lett.}\ }\textbf {\bibinfo {volume} {110}},\
  \bibinfo {pages} {171102} (\bibinfo {year} {2013})}\BibitemShut {NoStop}%
\bibitem [{\citenamefont {Graham}\ \emph
  {et~al.}(2016{\natexlab{a}})\citenamefont {Graham}, \citenamefont {Hogan},
  \citenamefont {Kasevich},\ and\ \citenamefont {Rajendran}}]{Graham2016_GW}%
  \BibitemOpen
  \bibfield  {author} {\bibinfo {author} {\bibfnamefont {P.~W.}\ \bibnamefont
  {Graham}}, \bibinfo {author} {\bibfnamefont {J.~M.}\ \bibnamefont {Hogan}},
  \bibinfo {author} {\bibfnamefont {M.~A.}\ \bibnamefont {Kasevich}},\ and\
  \bibinfo {author} {\bibfnamefont {S.}~\bibnamefont {Rajendran}},\ }\href
  {https://doi.org/10.1103/PhysRevD.94.104022} {\bibfield  {journal} {\bibinfo
  {journal} {Phys. Rev. D}\ }\textbf {\bibinfo {volume} {94}},\ \bibinfo
  {pages} {104022} (\bibinfo {year} {2016}{\natexlab{a}})}\BibitemShut
  {NoStop}%
\bibitem [{\citenamefont {Chaibi}\ \emph {et~al.}(2016)\citenamefont {Chaibi},
  \citenamefont {Geiger}, \citenamefont {Canuel}, \citenamefont {Bertoldi},
  \citenamefont {Landragin},\ and\ \citenamefont
  {Bouyer}}]{PhysRevD.93.021101}%
  \BibitemOpen
  \bibfield  {author} {\bibinfo {author} {\bibfnamefont {W.}~\bibnamefont
  {Chaibi}}, \bibinfo {author} {\bibfnamefont {R.}~\bibnamefont {Geiger}},
  \bibinfo {author} {\bibfnamefont {B.}~\bibnamefont {Canuel}}, \bibinfo
  {author} {\bibfnamefont {A.}~\bibnamefont {Bertoldi}}, \bibinfo {author}
  {\bibfnamefont {A.}~\bibnamefont {Landragin}},\ and\ \bibinfo {author}
  {\bibfnamefont {P.}~\bibnamefont {Bouyer}},\ }\href
  {https://doi.org/10.1103/PhysRevD.93.021101} {\bibfield  {journal} {\bibinfo
  {journal} {Phys. Rev. D}\ }\textbf {\bibinfo {volume} {93}},\ \bibinfo
  {pages} {021101} (\bibinfo {year} {2016})}\BibitemShut {NoStop}%
\bibitem [{\citenamefont {Canuel}\ \emph {et~al.}(2018)\citenamefont {Canuel},
  \citenamefont {Bertoldi}, \citenamefont {Amand}, \citenamefont {Pozzo~di
  Borgo}, \citenamefont {Chantrait}, \citenamefont {Danquigny}, \citenamefont
  {Dovale~{\'A}lvarez}, \citenamefont {Fang}, \citenamefont {Freise},
  \citenamefont {Geiger} \emph {et~al.}}]{canuel2018exploring}%
  \BibitemOpen
  \bibfield  {author} {\bibinfo {author} {\bibfnamefont {B.}~\bibnamefont
  {Canuel}}, \bibinfo {author} {\bibfnamefont {A.}~\bibnamefont {Bertoldi}},
  \bibinfo {author} {\bibfnamefont {L.}~\bibnamefont {Amand}}, \bibinfo
  {author} {\bibfnamefont {E.}~\bibnamefont {Pozzo~di Borgo}}, \bibinfo
  {author} {\bibfnamefont {T.}~\bibnamefont {Chantrait}}, \bibinfo {author}
  {\bibfnamefont {C.}~\bibnamefont {Danquigny}}, \bibinfo {author}
  {\bibfnamefont {M.}~\bibnamefont {Dovale~{\'A}lvarez}}, \bibinfo {author}
  {\bibfnamefont {B.}~\bibnamefont {Fang}}, \bibinfo {author} {\bibfnamefont
  {A.}~\bibnamefont {Freise}}, \bibinfo {author} {\bibfnamefont
  {R.}~\bibnamefont {Geiger}}, \emph {et~al.},\ }\href@noop {} {\bibfield
  {journal} {\bibinfo  {journal} {Sci. Rep.}\ }\textbf {\bibinfo {volume}
  {8}},\ \bibinfo {pages} {1} (\bibinfo {year} {2018})}\BibitemShut {NoStop}%
\bibitem [{\citenamefont {Hogan}\ \emph {et~al.}(2011)\citenamefont {Hogan},
  \citenamefont {Johnson}, \citenamefont {Dickerson}, \citenamefont {Kovachy},
  \citenamefont {Sugarbaker}, \citenamefont {Chiow}, \citenamefont {Graham},
  \citenamefont {Kasevich}, \citenamefont {Saif}, \citenamefont {Rajendran}
  \emph {et~al.}}]{hogan2011atomic}%
  \BibitemOpen
  \bibfield  {author} {\bibinfo {author} {\bibfnamefont {J.~M.}\ \bibnamefont
  {Hogan}}, \bibinfo {author} {\bibfnamefont {D.}~\bibnamefont {Johnson}},
  \bibinfo {author} {\bibfnamefont {S.}~\bibnamefont {Dickerson}}, \bibinfo
  {author} {\bibfnamefont {T.}~\bibnamefont {Kovachy}}, \bibinfo {author}
  {\bibfnamefont {A.}~\bibnamefont {Sugarbaker}}, \bibinfo {author}
  {\bibfnamefont {S.-w.}\ \bibnamefont {Chiow}}, \bibinfo {author}
  {\bibfnamefont {P.~W.}\ \bibnamefont {Graham}}, \bibinfo {author}
  {\bibfnamefont {M.~A.}\ \bibnamefont {Kasevich}}, \bibinfo {author}
  {\bibfnamefont {B.}~\bibnamefont {Saif}}, \bibinfo {author} {\bibfnamefont
  {S.}~\bibnamefont {Rajendran}}, \emph {et~al.},\ }\href@noop {} {\bibfield
  {journal} {\bibinfo  {journal} {Gen. Relativ. Gravit.}\ }\textbf {\bibinfo
  {volume} {43}},\ \bibinfo {pages} {1953} (\bibinfo {year}
  {2011})}\BibitemShut {NoStop}%
\bibitem [{\citenamefont {El-Neaj}\ \emph {et~al.}(2020)\citenamefont
  {El-Neaj}, \citenamefont {Alpigiani}, \citenamefont {Amairi-Pyka},
  \citenamefont {Ara{\'u}jo}, \citenamefont {Bala{\v{z}}}, \citenamefont
  {Bassi}, \citenamefont {Bathe-Peters}, \citenamefont {Battelier},
  \citenamefont {Beli{\'{c}}}, \citenamefont {Bentine}, \citenamefont
  {Bernabeu}, \citenamefont {Bertoldi}, \citenamefont {Bingham} \emph
  {et~al.}}]{abou2020aedge}%
  \BibitemOpen
  \bibfield  {author} {\bibinfo {author} {\bibfnamefont {Y.~A.}\ \bibnamefont
  {El-Neaj}}, \bibinfo {author} {\bibfnamefont {C.}~\bibnamefont {Alpigiani}},
  \bibinfo {author} {\bibfnamefont {S.}~\bibnamefont {Amairi-Pyka}}, \bibinfo
  {author} {\bibfnamefont {H.}~\bibnamefont {Ara{\'u}jo}}, \bibinfo {author}
  {\bibfnamefont {A.}~\bibnamefont {Bala{\v{z}}}}, \bibinfo {author}
  {\bibfnamefont {A.}~\bibnamefont {Bassi}}, \bibinfo {author} {\bibfnamefont
  {L.}~\bibnamefont {Bathe-Peters}}, \bibinfo {author} {\bibfnamefont
  {B.}~\bibnamefont {Battelier}}, \bibinfo {author} {\bibfnamefont
  {A.}~\bibnamefont {Beli{\'{c}}}}, \bibinfo {author} {\bibfnamefont
  {E.}~\bibnamefont {Bentine}}, \bibinfo {author} {\bibfnamefont
  {J.}~\bibnamefont {Bernabeu}}, \bibinfo {author} {\bibfnamefont
  {A.}~\bibnamefont {Bertoldi}}, \bibinfo {author} {\bibfnamefont
  {R.}~\bibnamefont {Bingham}}, \emph {et~al.},\ }\href
  {https://doi.org/10.1140/epjqt/s40507-020-0080-0} {\bibfield  {journal}
  {\bibinfo  {journal} {EPJ Quantum Technol.}\ }\textbf {\bibinfo {volume}
  {7}},\ \bibinfo {pages} {6} (\bibinfo {year} {2020})}\BibitemShut {NoStop}%
\bibitem [{\citenamefont {Badurina}\ \emph {et~al.}(2020)\citenamefont
  {Badurina}, \citenamefont {Bentine}, \citenamefont {Blas}, \citenamefont
  {Bongs}, \citenamefont {Bortoletto}, \citenamefont {Bowcock}, \citenamefont
  {Bridges}, \citenamefont {Bowden}, \citenamefont {Buchmueller}, \citenamefont
  {Burrage}, \citenamefont {Coleman}, \citenamefont {Elertas}, \citenamefont
  {Ellis}, \citenamefont {Foot}, \citenamefont {Gibson}, \citenamefont
  {Haehnelt}, \citenamefont {Harte}, \citenamefont {Hedges}, \citenamefont
  {Hobson}, \citenamefont {Holynski}, \citenamefont {Jones}, \citenamefont
  {Langlois}, \citenamefont {Lellouch}, \citenamefont {Lewicki}, \citenamefont
  {Maiolino}, \citenamefont {Majewski}, \citenamefont {Malik}, \citenamefont
  {March-Russell}, \citenamefont {McCabe}, \citenamefont {Newbold},
  \citenamefont {Sauer}, \citenamefont {Schneider}, \citenamefont {Shipsey},
  \citenamefont {Singh}, \citenamefont {Uchida}, \citenamefont {Valenzuela},
  \citenamefont {van~der Grinten}, \citenamefont {Vaskonen}, \citenamefont
  {Vossebeld}, \citenamefont {Weatherill},\ and\ \citenamefont
  {Wilmut}}]{Badurina_2020}%
  \BibitemOpen
  \bibfield  {author} {\bibinfo {author} {\bibfnamefont {L.}~\bibnamefont
  {Badurina}}, \bibinfo {author} {\bibfnamefont {E.}~\bibnamefont {Bentine}},
  \bibinfo {author} {\bibfnamefont {D.}~\bibnamefont {Blas}}, \bibinfo {author}
  {\bibfnamefont {K.}~\bibnamefont {Bongs}}, \bibinfo {author} {\bibfnamefont
  {D.}~\bibnamefont {Bortoletto}}, \bibinfo {author} {\bibfnamefont
  {T.}~\bibnamefont {Bowcock}}, \bibinfo {author} {\bibfnamefont
  {K.}~\bibnamefont {Bridges}}, \bibinfo {author} {\bibfnamefont
  {W.}~\bibnamefont {Bowden}}, \bibinfo {author} {\bibfnamefont
  {O.}~\bibnamefont {Buchmueller}}, \bibinfo {author} {\bibfnamefont
  {C.}~\bibnamefont {Burrage}}, \bibinfo {author} {\bibfnamefont
  {J.}~\bibnamefont {Coleman}}, \bibinfo {author} {\bibfnamefont
  {G.}~\bibnamefont {Elertas}}, \bibinfo {author} {\bibfnamefont
  {J.}~\bibnamefont {Ellis}}, \bibinfo {author} {\bibfnamefont
  {C.}~\bibnamefont {Foot}}, \bibinfo {author} {\bibfnamefont {V.}~\bibnamefont
  {Gibson}}, \bibinfo {author} {\bibfnamefont {M.}~\bibnamefont {Haehnelt}},
  \bibinfo {author} {\bibfnamefont {T.}~\bibnamefont {Harte}}, \bibinfo
  {author} {\bibfnamefont {S.}~\bibnamefont {Hedges}}, \bibinfo {author}
  {\bibfnamefont {R.}~\bibnamefont {Hobson}}, \bibinfo {author} {\bibfnamefont
  {M.}~\bibnamefont {Holynski}}, \bibinfo {author} {\bibfnamefont
  {T.}~\bibnamefont {Jones}}, \bibinfo {author} {\bibfnamefont
  {M.}~\bibnamefont {Langlois}}, \bibinfo {author} {\bibfnamefont
  {S.}~\bibnamefont {Lellouch}}, \bibinfo {author} {\bibfnamefont
  {M.}~\bibnamefont {Lewicki}}, \bibinfo {author} {\bibfnamefont
  {R.}~\bibnamefont {Maiolino}}, \bibinfo {author} {\bibfnamefont
  {P.}~\bibnamefont {Majewski}}, \bibinfo {author} {\bibfnamefont
  {S.}~\bibnamefont {Malik}}, \bibinfo {author} {\bibfnamefont
  {J.}~\bibnamefont {March-Russell}}, \bibinfo {author} {\bibfnamefont
  {C.}~\bibnamefont {McCabe}}, \bibinfo {author} {\bibfnamefont
  {D.}~\bibnamefont {Newbold}}, \bibinfo {author} {\bibfnamefont
  {B.}~\bibnamefont {Sauer}}, \bibinfo {author} {\bibfnamefont
  {U.}~\bibnamefont {Schneider}}, \bibinfo {author} {\bibfnamefont
  {I.}~\bibnamefont {Shipsey}}, \bibinfo {author} {\bibfnamefont
  {Y.}~\bibnamefont {Singh}}, \bibinfo {author} {\bibfnamefont
  {M.}~\bibnamefont {Uchida}}, \bibinfo {author} {\bibfnamefont
  {T.}~\bibnamefont {Valenzuela}}, \bibinfo {author} {\bibfnamefont
  {M.}~\bibnamefont {van~der Grinten}}, \bibinfo {author} {\bibfnamefont
  {V.}~\bibnamefont {Vaskonen}}, \bibinfo {author} {\bibfnamefont
  {J.}~\bibnamefont {Vossebeld}}, \bibinfo {author} {\bibfnamefont
  {D.}~\bibnamefont {Weatherill}},\ and\ \bibinfo {author} {\bibfnamefont
  {I.}~\bibnamefont {Wilmut}},\ }\href
  {https://doi.org/10.1088/1475-7516/2020/05/011} {\bibfield  {journal}
  {\bibinfo  {journal} {J. Cosmol. Astropart. Phys.}\ }\textbf {\bibinfo
  {volume} {2020}}\bibinfo  {number} { (05)},\ \bibinfo {pages}
  {011}}\BibitemShut {NoStop}%
\bibitem [{\citenamefont {Zhan}\ \emph {et~al.}(2020)\citenamefont {Zhan},
  \citenamefont {Wang}, \citenamefont {Ni}, \citenamefont {Gao}, \citenamefont
  {Wang}, \citenamefont {He}, \citenamefont {Li}, \citenamefont {Zhou},
  \citenamefont {Chen}, \citenamefont {Zhong}, \citenamefont {Tang},
  \citenamefont {Yao}, \citenamefont {Zhu}, \citenamefont {Xiong},
  \citenamefont {Lu}, \citenamefont {Yu}, \citenamefont {Cheng}, \citenamefont
  {Liu}, \citenamefont {Liang}, \citenamefont {Xu}, \citenamefont {He},
  \citenamefont {Ke}, \citenamefont {Tan},\ and\ \citenamefont
  {Luo}}]{ZAIGA2020}%
  \BibitemOpen
\bibfield  {number} {  }\bibfield  {author} {\bibinfo {author} {\bibfnamefont
  {M.-S.}\ \bibnamefont {Zhan}}, \bibinfo {author} {\bibfnamefont
  {J.}~\bibnamefont {Wang}}, \bibinfo {author} {\bibfnamefont {W.-T.}\
  \bibnamefont {Ni}}, \bibinfo {author} {\bibfnamefont {D.-F.}\ \bibnamefont
  {Gao}}, \bibinfo {author} {\bibfnamefont {G.}~\bibnamefont {Wang}}, \bibinfo
  {author} {\bibfnamefont {L.-X.}\ \bibnamefont {He}}, \bibinfo {author}
  {\bibfnamefont {R.-B.}\ \bibnamefont {Li}}, \bibinfo {author} {\bibfnamefont
  {L.}~\bibnamefont {Zhou}}, \bibinfo {author} {\bibfnamefont {X.}~\bibnamefont
  {Chen}}, \bibinfo {author} {\bibfnamefont {J.-Q.}\ \bibnamefont {Zhong}},
  \bibinfo {author} {\bibfnamefont {B.}~\bibnamefont {Tang}}, \bibinfo {author}
  {\bibfnamefont {Z.-W.}\ \bibnamefont {Yao}}, \bibinfo {author} {\bibfnamefont
  {L.}~\bibnamefont {Zhu}}, \bibinfo {author} {\bibfnamefont {Z.-Y.}\
  \bibnamefont {Xiong}}, \bibinfo {author} {\bibfnamefont {S.-B.}\ \bibnamefont
  {Lu}}, \bibinfo {author} {\bibfnamefont {G.-H.}\ \bibnamefont {Yu}}, \bibinfo
  {author} {\bibfnamefont {Q.-F.}\ \bibnamefont {Cheng}}, \bibinfo {author}
  {\bibfnamefont {M.}~\bibnamefont {Liu}}, \bibinfo {author} {\bibfnamefont
  {Y.-R.}\ \bibnamefont {Liang}}, \bibinfo {author} {\bibfnamefont
  {P.}~\bibnamefont {Xu}}, \bibinfo {author} {\bibfnamefont {X.-D.}\
  \bibnamefont {He}}, \bibinfo {author} {\bibfnamefont {M.}~\bibnamefont {Ke}},
  \bibinfo {author} {\bibfnamefont {Z.}~\bibnamefont {Tan}},\ and\ \bibinfo
  {author} {\bibfnamefont {J.}~\bibnamefont {Luo}},\ }\href
  {https://doi.org/10.1142/S0218271819400054} {\bibfield  {journal} {\bibinfo
  {journal} {Int. J. Mod. Phys. D}\ }\textbf {\bibinfo {volume} {29}},\
  \bibinfo {pages} {1940005} (\bibinfo {year} {2020})}\BibitemShut {NoStop}%
\bibitem [{\citenamefont {Abe}\ \emph {et~al.}(2021)\citenamefont {Abe},
  \citenamefont {Adamson}, \citenamefont {Borcean}, \citenamefont {Bortoletto},
  \citenamefont {Bridges}, \citenamefont {Carman}, \citenamefont
  {Chattopadhyay}, \citenamefont {Coleman}, \citenamefont {Curfman},
  \citenamefont {DeRose} \emph {et~al.}}]{abe2021matter}%
  \BibitemOpen
  \bibfield  {author} {\bibinfo {author} {\bibfnamefont {M.}~\bibnamefont
  {Abe}}, \bibinfo {author} {\bibfnamefont {P.}~\bibnamefont {Adamson}},
  \bibinfo {author} {\bibfnamefont {M.}~\bibnamefont {Borcean}}, \bibinfo
  {author} {\bibfnamefont {D.}~\bibnamefont {Bortoletto}}, \bibinfo {author}
  {\bibfnamefont {K.}~\bibnamefont {Bridges}}, \bibinfo {author} {\bibfnamefont
  {S.~P.}\ \bibnamefont {Carman}}, \bibinfo {author} {\bibfnamefont
  {S.}~\bibnamefont {Chattopadhyay}}, \bibinfo {author} {\bibfnamefont
  {J.}~\bibnamefont {Coleman}}, \bibinfo {author} {\bibfnamefont {N.~M.}\
  \bibnamefont {Curfman}}, \bibinfo {author} {\bibfnamefont {K.}~\bibnamefont
  {DeRose}}, \emph {et~al.},\ }\href@noop {} {\bibfield  {journal} {\bibinfo
  {journal} {Quantum Sci. Technol.}\ }\textbf {\bibinfo {volume} {6}},\
  \bibinfo {pages} {044003} (\bibinfo {year} {2021})}\BibitemShut {NoStop}%
\bibitem [{\citenamefont {Bouchendira}\ \emph {et~al.}(2011)\citenamefont
  {Bouchendira}, \citenamefont {Clad\'e}, \citenamefont {Guellati-Kh\'elifa},
  \citenamefont {Nez},\ and\ \citenamefont
  {Biraben}}]{Bouchendira2011_finestructure}%
  \BibitemOpen
  \bibfield  {author} {\bibinfo {author} {\bibfnamefont {R.}~\bibnamefont
  {Bouchendira}}, \bibinfo {author} {\bibfnamefont {P.}~\bibnamefont
  {Clad\'e}}, \bibinfo {author} {\bibfnamefont {S.}~\bibnamefont
  {Guellati-Kh\'elifa}}, \bibinfo {author} {\bibfnamefont {F.~m.~c.}\
  \bibnamefont {Nez}},\ and\ \bibinfo {author} {\bibfnamefont {F.~m.~c.}\
  \bibnamefont {Biraben}},\ }\href
  {https://doi.org/10.1103/PhysRevLett.106.080801} {\bibfield  {journal}
  {\bibinfo  {journal} {Phys. Rev. Lett.}\ }\textbf {\bibinfo {volume} {106}},\
  \bibinfo {pages} {080801} (\bibinfo {year} {2011})}\BibitemShut {NoStop}%
\bibitem [{\citenamefont {Parker}\ \emph {et~al.}(2018)\citenamefont {Parker},
  \citenamefont {Yu}, \citenamefont {Zhong}, \citenamefont {Estey},\ and\
  \citenamefont {M{\"u}ller}}]{parker2018finestructure}%
  \BibitemOpen
  \bibfield  {author} {\bibinfo {author} {\bibfnamefont {R.~H.}\ \bibnamefont
  {Parker}}, \bibinfo {author} {\bibfnamefont {C.}~\bibnamefont {Yu}}, \bibinfo
  {author} {\bibfnamefont {W.}~\bibnamefont {Zhong}}, \bibinfo {author}
  {\bibfnamefont {B.}~\bibnamefont {Estey}},\ and\ \bibinfo {author}
  {\bibfnamefont {H.}~\bibnamefont {M{\"u}ller}},\ }\href@noop {} {\bibfield
  {journal} {\bibinfo  {journal} {Science}\ }\textbf {\bibinfo {volume}
  {360}},\ \bibinfo {pages} {191} (\bibinfo {year} {2018})}\BibitemShut
  {NoStop}%
\bibitem [{\citenamefont {Morel}\ \emph {et~al.}(2020)\citenamefont {Morel},
  \citenamefont {Yao}, \citenamefont {Clad{\'e}},\ and\ \citenamefont
  {Guellati-Kh{\'e}lifa}}]{morel2020determination}%
  \BibitemOpen
  \bibfield  {author} {\bibinfo {author} {\bibfnamefont {L.}~\bibnamefont
  {Morel}}, \bibinfo {author} {\bibfnamefont {Z.}~\bibnamefont {Yao}}, \bibinfo
  {author} {\bibfnamefont {P.}~\bibnamefont {Clad{\'e}}},\ and\ \bibinfo
  {author} {\bibfnamefont {S.}~\bibnamefont {Guellati-Kh{\'e}lifa}},\
  }\href@noop {} {\bibfield  {journal} {\bibinfo  {journal} {Nature}\ }\textbf
  {\bibinfo {volume} {588}},\ \bibinfo {pages} {61} (\bibinfo {year}
  {2020})}\BibitemShut {NoStop}%
\bibitem [{\citenamefont {Biedermann}\ \emph {et~al.}(2015)\citenamefont
  {Biedermann}, \citenamefont {Wu}, \citenamefont {Deslauriers}, \citenamefont
  {Roy}, \citenamefont {Mahadeswaraswamy},\ and\ \citenamefont
  {Kasevich}}]{Biedermann2015_gravity}%
  \BibitemOpen
  \bibfield  {author} {\bibinfo {author} {\bibfnamefont {G.~W.}\ \bibnamefont
  {Biedermann}}, \bibinfo {author} {\bibfnamefont {X.}~\bibnamefont {Wu}},
  \bibinfo {author} {\bibfnamefont {L.}~\bibnamefont {Deslauriers}}, \bibinfo
  {author} {\bibfnamefont {S.}~\bibnamefont {Roy}}, \bibinfo {author}
  {\bibfnamefont {C.}~\bibnamefont {Mahadeswaraswamy}},\ and\ \bibinfo {author}
  {\bibfnamefont {M.~A.}\ \bibnamefont {Kasevich}},\ }\href
  {https://doi.org/10.1103/PhysRevA.91.033629} {\bibfield  {journal} {\bibinfo
  {journal} {Phys. Rev. A}\ }\textbf {\bibinfo {volume} {91}},\ \bibinfo
  {pages} {033629} (\bibinfo {year} {2015})}\BibitemShut {NoStop}%
\bibitem [{\citenamefont {Rosi}\ \emph {et~al.}(2014)\citenamefont {Rosi},
  \citenamefont {Sorrentino}, \citenamefont {Cacciapuoti}, \citenamefont
  {Prevedelli},\ and\ \citenamefont {Tino}}]{rosi2014precision}%
  \BibitemOpen
  \bibfield  {author} {\bibinfo {author} {\bibfnamefont {G.}~\bibnamefont
  {Rosi}}, \bibinfo {author} {\bibfnamefont {F.}~\bibnamefont {Sorrentino}},
  \bibinfo {author} {\bibfnamefont {L.}~\bibnamefont {Cacciapuoti}}, \bibinfo
  {author} {\bibfnamefont {M.}~\bibnamefont {Prevedelli}},\ and\ \bibinfo
  {author} {\bibfnamefont {G.}~\bibnamefont {Tino}},\ }\href@noop {} {\bibfield
   {journal} {\bibinfo  {journal} {Nature}\ }\textbf {\bibinfo {volume}
  {510}},\ \bibinfo {pages} {518} (\bibinfo {year} {2014})}\BibitemShut
  {NoStop}%
\bibitem [{\citenamefont {Arvanitaki}\ \emph {et~al.}(2018)\citenamefont
  {Arvanitaki}, \citenamefont {Graham}, \citenamefont {Hogan}, \citenamefont
  {Rajendran},\ and\ \citenamefont {Van~Tilburg}}]{Arvanitaki2018_darkmatter}%
  \BibitemOpen
  \bibfield  {author} {\bibinfo {author} {\bibfnamefont {A.}~\bibnamefont
  {Arvanitaki}}, \bibinfo {author} {\bibfnamefont {P.~W.}\ \bibnamefont
  {Graham}}, \bibinfo {author} {\bibfnamefont {J.~M.}\ \bibnamefont {Hogan}},
  \bibinfo {author} {\bibfnamefont {S.}~\bibnamefont {Rajendran}},\ and\
  \bibinfo {author} {\bibfnamefont {K.}~\bibnamefont {Van~Tilburg}},\ }\href
  {https://doi.org/10.1103/PhysRevD.97.075020} {\bibfield  {journal} {\bibinfo
  {journal} {Phys. Rev. D}\ }\textbf {\bibinfo {volume} {97}},\ \bibinfo
  {pages} {075020} (\bibinfo {year} {2018})}\BibitemShut {NoStop}%
\bibitem [{\citenamefont {Graham}\ \emph
  {et~al.}(2016{\natexlab{b}})\citenamefont {Graham}, \citenamefont {Kaplan},
  \citenamefont {Mardon}, \citenamefont {Rajendran},\ and\ \citenamefont
  {Terrano}}]{Graham2016_darkmatter}%
  \BibitemOpen
  \bibfield  {author} {\bibinfo {author} {\bibfnamefont {P.~W.}\ \bibnamefont
  {Graham}}, \bibinfo {author} {\bibfnamefont {D.~E.}\ \bibnamefont {Kaplan}},
  \bibinfo {author} {\bibfnamefont {J.}~\bibnamefont {Mardon}}, \bibinfo
  {author} {\bibfnamefont {S.}~\bibnamefont {Rajendran}},\ and\ \bibinfo
  {author} {\bibfnamefont {W.~A.}\ \bibnamefont {Terrano}},\ }\href
  {https://doi.org/10.1103/PhysRevD.93.075029} {\bibfield  {journal} {\bibinfo
  {journal} {Phys. Rev. D}\ }\textbf {\bibinfo {volume} {93}},\ \bibinfo
  {pages} {075029} (\bibinfo {year} {2016}{\natexlab{b}})}\BibitemShut
  {NoStop}%
\bibitem [{\citenamefont {Banerjee}\ \emph {et~al.}(2022)\citenamefont
  {Banerjee}, \citenamefont {Perez}, \citenamefont {Safronova}, \citenamefont
  {Savoray},\ and\ \citenamefont {Shalit}}]{banerjee2022phenomenology}%
  \BibitemOpen
  \bibfield  {author} {\bibinfo {author} {\bibfnamefont {A.}~\bibnamefont
  {Banerjee}}, \bibinfo {author} {\bibfnamefont {G.}~\bibnamefont {Perez}},
  \bibinfo {author} {\bibfnamefont {M.}~\bibnamefont {Safronova}}, \bibinfo
  {author} {\bibfnamefont {I.}~\bibnamefont {Savoray}},\ and\ \bibinfo {author}
  {\bibfnamefont {A.}~\bibnamefont {Shalit}},\ }\href@noop {} {\bibfield
  {journal} {\bibinfo  {journal} {arXiv preprint arXiv:2211.05174}\ } (\bibinfo
  {year} {2022})}\BibitemShut {NoStop}%
\bibitem [{\citenamefont {Hamilton}\ \emph {et~al.}(2015)\citenamefont
  {Hamilton}, \citenamefont {Jaffe}, \citenamefont {Haslinger}, \citenamefont
  {Simmons}, \citenamefont {M{\"u}ller},\ and\ \citenamefont
  {Khoury}}]{hamilton2015_darkenergy}%
  \BibitemOpen
  \bibfield  {author} {\bibinfo {author} {\bibfnamefont {P.}~\bibnamefont
  {Hamilton}}, \bibinfo {author} {\bibfnamefont {M.}~\bibnamefont {Jaffe}},
  \bibinfo {author} {\bibfnamefont {P.}~\bibnamefont {Haslinger}}, \bibinfo
  {author} {\bibfnamefont {Q.}~\bibnamefont {Simmons}}, \bibinfo {author}
  {\bibfnamefont {H.}~\bibnamefont {M{\"u}ller}},\ and\ \bibinfo {author}
  {\bibfnamefont {J.}~\bibnamefont {Khoury}},\ }\href@noop {} {\bibfield
  {journal} {\bibinfo  {journal} {Science}\ }\textbf {\bibinfo {volume}
  {349}},\ \bibinfo {pages} {849} (\bibinfo {year} {2015})}\BibitemShut
  {NoStop}%
\bibitem [{\citenamefont {Wu}\ \emph {et~al.}(2019)\citenamefont {Wu},
  \citenamefont {Pagel}, \citenamefont {Malek}, \citenamefont {Nguyen},
  \citenamefont {Zi}, \citenamefont {Scheirer},\ and\ \citenamefont
  {Müller}}]{Wu2019_mobile}%
  \BibitemOpen
  \bibfield  {author} {\bibinfo {author} {\bibfnamefont {X.}~\bibnamefont
  {Wu}}, \bibinfo {author} {\bibfnamefont {Z.}~\bibnamefont {Pagel}}, \bibinfo
  {author} {\bibfnamefont {B.~S.}\ \bibnamefont {Malek}}, \bibinfo {author}
  {\bibfnamefont {T.~H.}\ \bibnamefont {Nguyen}}, \bibinfo {author}
  {\bibfnamefont {F.}~\bibnamefont {Zi}}, \bibinfo {author} {\bibfnamefont
  {D.~S.}\ \bibnamefont {Scheirer}},\ and\ \bibinfo {author} {\bibfnamefont
  {H.}~\bibnamefont {Müller}},\ }\href@noop {} {\bibfield  {journal} {\bibinfo
   {journal} {Sci. Adv.}\ }\textbf {\bibinfo {volume} {5}},\ \bibinfo {pages}
  {eaax0800} (\bibinfo {year} {2019})}\BibitemShut {NoStop}%
\bibitem [{\citenamefont {Bongs}\ \emph {et~al.}(2019)\citenamefont {Bongs},
  \citenamefont {Holynski}, \citenamefont {Vovrosh}, \citenamefont {Bouyer},
  \citenamefont {Condon}, \citenamefont {Rasel}, \citenamefont {Schubert},
  \citenamefont {Schleich},\ and\ \citenamefont {Roura}}]{bongs2019taking}%
  \BibitemOpen
  \bibfield  {author} {\bibinfo {author} {\bibfnamefont {K.}~\bibnamefont
  {Bongs}}, \bibinfo {author} {\bibfnamefont {M.}~\bibnamefont {Holynski}},
  \bibinfo {author} {\bibfnamefont {J.}~\bibnamefont {Vovrosh}}, \bibinfo
  {author} {\bibfnamefont {P.}~\bibnamefont {Bouyer}}, \bibinfo {author}
  {\bibfnamefont {G.}~\bibnamefont {Condon}}, \bibinfo {author} {\bibfnamefont
  {E.}~\bibnamefont {Rasel}}, \bibinfo {author} {\bibfnamefont
  {C.}~\bibnamefont {Schubert}}, \bibinfo {author} {\bibfnamefont {W.~P.}\
  \bibnamefont {Schleich}},\ and\ \bibinfo {author} {\bibfnamefont
  {A.}~\bibnamefont {Roura}},\ }\href
  {https://doi.org/10.1038/s42254-019-0117-4} {\bibfield  {journal} {\bibinfo
  {journal} {Nat. Rev. Phys.}\ }\textbf {\bibinfo {volume} {1}},\ \bibinfo
  {pages} {731} (\bibinfo {year} {2019})}\BibitemShut {NoStop}%
\bibitem [{\citenamefont {McGuirk}\ \emph {et~al.}(2000)\citenamefont
  {McGuirk}, \citenamefont {Snadden},\ and\ \citenamefont
  {Kasevich}}]{mcguirk2000}%
  \BibitemOpen
  \bibfield  {author} {\bibinfo {author} {\bibfnamefont {J.~M.}\ \bibnamefont
  {McGuirk}}, \bibinfo {author} {\bibfnamefont {M.~J.}\ \bibnamefont
  {Snadden}},\ and\ \bibinfo {author} {\bibfnamefont {M.~A.}\ \bibnamefont
  {Kasevich}},\ }\href {https://doi.org/10.1103/PhysRevLett.85.4498} {\bibfield
   {journal} {\bibinfo  {journal} {Phys. Rev. Lett.}\ }\textbf {\bibinfo
  {volume} {85}},\ \bibinfo {pages} {4498} (\bibinfo {year}
  {2000})}\BibitemShut {NoStop}%
\bibitem [{\citenamefont {M\"uller}\ \emph {et~al.}(2008)\citenamefont
  {M\"uller}, \citenamefont {Chiow}, \citenamefont {Long}, \citenamefont
  {Herrmann},\ and\ \citenamefont {Chu}}]{muller2008}%
  \BibitemOpen
  \bibfield  {author} {\bibinfo {author} {\bibfnamefont {H.}~\bibnamefont
  {M\"uller}}, \bibinfo {author} {\bibfnamefont {S.-w.}\ \bibnamefont {Chiow}},
  \bibinfo {author} {\bibfnamefont {Q.}~\bibnamefont {Long}}, \bibinfo {author}
  {\bibfnamefont {S.}~\bibnamefont {Herrmann}},\ and\ \bibinfo {author}
  {\bibfnamefont {S.}~\bibnamefont {Chu}},\ }\href
  {https://doi.org/10.1103/PhysRevLett.100.180405} {\bibfield  {journal}
  {\bibinfo  {journal} {Phys. Rev. Lett.}\ }\textbf {\bibinfo {volume} {100}},\
  \bibinfo {pages} {180405} (\bibinfo {year} {2008})}\BibitemShut {NoStop}%
\bibitem [{\citenamefont {M\"uller}\ \emph {et~al.}(2009)\citenamefont
  {M\"uller}, \citenamefont {Chiow}, \citenamefont {Herrmann},\ and\
  \citenamefont {Chu}}]{muller2009}%
  \BibitemOpen
  \bibfield  {author} {\bibinfo {author} {\bibfnamefont {H.}~\bibnamefont
  {M\"uller}}, \bibinfo {author} {\bibfnamefont {S.-w.}\ \bibnamefont {Chiow}},
  \bibinfo {author} {\bibfnamefont {S.}~\bibnamefont {Herrmann}},\ and\
  \bibinfo {author} {\bibfnamefont {S.}~\bibnamefont {Chu}},\ }\href
  {https://doi.org/10.1103/PhysRevLett.102.240403} {\bibfield  {journal}
  {\bibinfo  {journal} {Phys. Rev. Lett.}\ }\textbf {\bibinfo {volume} {102}},\
  \bibinfo {pages} {240403} (\bibinfo {year} {2009})}\BibitemShut {NoStop}%
\bibitem [{\citenamefont {Clad\'e}\ \emph {et~al.}(2009)\citenamefont
  {Clad\'e}, \citenamefont {Guellati-Kh\'elifa}, \citenamefont {Nez},\ and\
  \citenamefont {Biraben}}]{clade2009}%
  \BibitemOpen
  \bibfield  {author} {\bibinfo {author} {\bibfnamefont {P.}~\bibnamefont
  {Clad\'e}}, \bibinfo {author} {\bibfnamefont {S.}~\bibnamefont
  {Guellati-Kh\'elifa}}, \bibinfo {author} {\bibfnamefont {F.~m.~c.}\
  \bibnamefont {Nez}},\ and\ \bibinfo {author} {\bibfnamefont {F.~m.~c.}\
  \bibnamefont {Biraben}},\ }\href
  {https://doi.org/10.1103/PhysRevLett.102.240402} {\bibfield  {journal}
  {\bibinfo  {journal} {Phys. Rev. Lett.}\ }\textbf {\bibinfo {volume} {102}},\
  \bibinfo {pages} {240402} (\bibinfo {year} {2009})}\BibitemShut {NoStop}%
\bibitem [{\citenamefont {Gebbe}\ \emph {et~al.}(2021)\citenamefont {Gebbe},
  \citenamefont {Siem{\ss}}, \citenamefont {Gersemann}, \citenamefont
  {M{\"u}ntinga}, \citenamefont {Herrmann}, \citenamefont {L{\"a}mmerzahl},
  \citenamefont {Ahlers}, \citenamefont {Gaaloul}, \citenamefont {Schubert},
  \citenamefont {Hammerer} \emph {et~al.}}]{gebbe2021twin}%
  \BibitemOpen
  \bibfield  {author} {\bibinfo {author} {\bibfnamefont {M.}~\bibnamefont
  {Gebbe}}, \bibinfo {author} {\bibfnamefont {J.-N.}\ \bibnamefont
  {Siem{\ss}}}, \bibinfo {author} {\bibfnamefont {M.}~\bibnamefont
  {Gersemann}}, \bibinfo {author} {\bibfnamefont {H.}~\bibnamefont
  {M{\"u}ntinga}}, \bibinfo {author} {\bibfnamefont {S.}~\bibnamefont
  {Herrmann}}, \bibinfo {author} {\bibfnamefont {C.}~\bibnamefont
  {L{\"a}mmerzahl}}, \bibinfo {author} {\bibfnamefont {H.}~\bibnamefont
  {Ahlers}}, \bibinfo {author} {\bibfnamefont {N.}~\bibnamefont {Gaaloul}},
  \bibinfo {author} {\bibfnamefont {C.}~\bibnamefont {Schubert}}, \bibinfo
  {author} {\bibfnamefont {K.}~\bibnamefont {Hammerer}}, \emph {et~al.},\
  }\href@noop {} {\bibfield  {journal} {\bibinfo  {journal} {Nat. Comm.}\
  }\textbf {\bibinfo {volume} {12}},\ \bibinfo {pages} {1} (\bibinfo {year}
  {2021})}\BibitemShut {NoStop}%
\bibitem [{\citenamefont {Chiow}\ \emph {et~al.}(2011)\citenamefont {Chiow},
  \citenamefont {Kovachy}, \citenamefont {Chien},\ and\ \citenamefont
  {Kasevich}}]{102hk_large_area}%
  \BibitemOpen
  \bibfield  {author} {\bibinfo {author} {\bibfnamefont {S.-w.}\ \bibnamefont
  {Chiow}}, \bibinfo {author} {\bibfnamefont {T.}~\bibnamefont {Kovachy}},
  \bibinfo {author} {\bibfnamefont {H.-C.}\ \bibnamefont {Chien}},\ and\
  \bibinfo {author} {\bibfnamefont {M.~A.}\ \bibnamefont {Kasevich}},\ }\href
  {https://doi.org/10.1103/PhysRevLett.107.130403} {\bibfield  {journal}
  {\bibinfo  {journal} {Phys. Rev. Lett.}\ }\textbf {\bibinfo {volume} {107}},\
  \bibinfo {pages} {130403} (\bibinfo {year} {2011})}\BibitemShut {NoStop}%
\bibitem [{\citenamefont {McDonald}\ \emph {et~al.}(2013)\citenamefont
  {McDonald}, \citenamefont {Kuhn}, \citenamefont {Bennetts}, \citenamefont
  {Debs}, \citenamefont {Hardman}, \citenamefont {Johnsson}, \citenamefont
  {Close},\ and\ \citenamefont {Robins}}]{Close:2013}%
  \BibitemOpen
  \bibfield  {author} {\bibinfo {author} {\bibfnamefont {G.~D.}\ \bibnamefont
  {McDonald}}, \bibinfo {author} {\bibfnamefont {C.~C.~N.}\ \bibnamefont
  {Kuhn}}, \bibinfo {author} {\bibfnamefont {S.}~\bibnamefont {Bennetts}},
  \bibinfo {author} {\bibfnamefont {J.~E.}\ \bibnamefont {Debs}}, \bibinfo
  {author} {\bibfnamefont {K.~S.}\ \bibnamefont {Hardman}}, \bibinfo {author}
  {\bibfnamefont {M.}~\bibnamefont {Johnsson}}, \bibinfo {author}
  {\bibfnamefont {J.~D.}\ \bibnamefont {Close}},\ and\ \bibinfo {author}
  {\bibfnamefont {N.~P.}\ \bibnamefont {Robins}},\ }\href
  {https://doi.org/10.1103/PhysRevA.88.053620} {\bibfield  {journal} {\bibinfo
  {journal} {Phys. Rev. A}\ }\textbf {\bibinfo {volume} {88}},\ \bibinfo
  {pages} {053620} (\bibinfo {year} {2013})}\BibitemShut {NoStop}%
\bibitem [{\citenamefont {Mazzoni}\ \emph {et~al.}(2015)\citenamefont
  {Mazzoni}, \citenamefont {Zhang}, \citenamefont {Del~Aguila}, \citenamefont
  {Salvi}, \citenamefont {Poli},\ and\ \citenamefont {Tino}}]{Mazzoni2015}%
  \BibitemOpen
  \bibfield  {author} {\bibinfo {author} {\bibfnamefont {T.}~\bibnamefont
  {Mazzoni}}, \bibinfo {author} {\bibfnamefont {X.}~\bibnamefont {Zhang}},
  \bibinfo {author} {\bibfnamefont {R.}~\bibnamefont {Del~Aguila}}, \bibinfo
  {author} {\bibfnamefont {L.}~\bibnamefont {Salvi}}, \bibinfo {author}
  {\bibfnamefont {N.}~\bibnamefont {Poli}},\ and\ \bibinfo {author}
  {\bibfnamefont {G.~M.}\ \bibnamefont {Tino}},\ }\href
  {https://doi.org/10.1103/PhysRevA.92.053619} {\bibfield  {journal} {\bibinfo
  {journal} {Phys. Rev. A}\ }\textbf {\bibinfo {volume} {92}},\ \bibinfo
  {pages} {053619} (\bibinfo {year} {2015})}\BibitemShut {NoStop}%
\bibitem [{\citenamefont {Kotru}\ \emph {et~al.}(2015)\citenamefont {Kotru},
  \citenamefont {Butts}, \citenamefont {Kinast},\ and\ \citenamefont
  {Stoner}}]{Kotru2015}%
  \BibitemOpen
  \bibfield  {author} {\bibinfo {author} {\bibfnamefont {K.}~\bibnamefont
  {Kotru}}, \bibinfo {author} {\bibfnamefont {D.~L.}\ \bibnamefont {Butts}},
  \bibinfo {author} {\bibfnamefont {J.~M.}\ \bibnamefont {Kinast}},\ and\
  \bibinfo {author} {\bibfnamefont {R.~E.}\ \bibnamefont {Stoner}},\ }\href
  {https://doi.org/10.1103/PhysRevLett.115.103001} {\bibfield  {journal}
  {\bibinfo  {journal} {Phys. Rev. Lett.}\ }\textbf {\bibinfo {volume} {115}},\
  \bibinfo {pages} {103001} (\bibinfo {year} {2015})}\BibitemShut {NoStop}%
\bibitem [{\citenamefont {Plotkin-Swing}\ \emph {et~al.}(2018)\citenamefont
  {Plotkin-Swing}, \citenamefont {Gochnauer}, \citenamefont {McAlpine},
  \citenamefont {Cooper}, \citenamefont {Jamison},\ and\ \citenamefont
  {Gupta}}]{Plotkin2018}%
  \BibitemOpen
  \bibfield  {author} {\bibinfo {author} {\bibfnamefont {B.}~\bibnamefont
  {Plotkin-Swing}}, \bibinfo {author} {\bibfnamefont {D.}~\bibnamefont
  {Gochnauer}}, \bibinfo {author} {\bibfnamefont {K.~E.}\ \bibnamefont
  {McAlpine}}, \bibinfo {author} {\bibfnamefont {E.~S.}\ \bibnamefont
  {Cooper}}, \bibinfo {author} {\bibfnamefont {A.~O.}\ \bibnamefont
  {Jamison}},\ and\ \bibinfo {author} {\bibfnamefont {S.}~\bibnamefont
  {Gupta}},\ }\href {https://doi.org/10.1103/PhysRevLett.121.133201} {\bibfield
   {journal} {\bibinfo  {journal} {Phys. Rev. Lett.}\ }\textbf {\bibinfo
  {volume} {121}},\ \bibinfo {pages} {133201} (\bibinfo {year}
  {2018})}\BibitemShut {NoStop}%
\bibitem [{\citenamefont {Pagel}\ \emph {et~al.}(2020)\citenamefont {Pagel},
  \citenamefont {Zhong}, \citenamefont {Parker}, \citenamefont {Olund},
  \citenamefont {Yao},\ and\ \citenamefont {M\"uller}}]{pagel2019bloch}%
  \BibitemOpen
  \bibfield  {author} {\bibinfo {author} {\bibfnamefont {Z.}~\bibnamefont
  {Pagel}}, \bibinfo {author} {\bibfnamefont {W.}~\bibnamefont {Zhong}},
  \bibinfo {author} {\bibfnamefont {R.~H.}\ \bibnamefont {Parker}}, \bibinfo
  {author} {\bibfnamefont {C.~T.}\ \bibnamefont {Olund}}, \bibinfo {author}
  {\bibfnamefont {N.~Y.}\ \bibnamefont {Yao}},\ and\ \bibinfo {author}
  {\bibfnamefont {H.}~\bibnamefont {M\"uller}},\ }\href
  {https://doi.org/10.1103/PhysRevA.102.053312} {\bibfield  {journal} {\bibinfo
   {journal} {Phys. Rev. A}\ }\textbf {\bibinfo {volume} {102}},\ \bibinfo
  {pages} {053312} (\bibinfo {year} {2020})}\BibitemShut {NoStop}%
\bibitem [{\citenamefont {Young}(1997)}]{young1997measurement}%
  \BibitemOpen
  \bibfield  {author} {\bibinfo {author} {\bibfnamefont {B.~C.}\ \bibnamefont
  {Young}},\ }\emph {\bibinfo {title} {A measurement of the fine-structure
  constant using atom interferometry}},\ \href@noop {} {Ph.D. thesis} (\bibinfo
  {year} {1997})\BibitemShut {NoStop}%
\bibitem [{\citenamefont {Hinkley}\ \emph {et~al.}(2013)\citenamefont
  {Hinkley}, \citenamefont {Sherman}, \citenamefont {Phillips}, \citenamefont
  {Schioppo}, \citenamefont {Lemke}, \citenamefont {Beloy}, \citenamefont
  {Pizzocaro}, \citenamefont {Oates},\ and\ \citenamefont
  {Ludlow}}]{hinkley2013clock}%
  \BibitemOpen
  \bibfield  {author} {\bibinfo {author} {\bibfnamefont {N.}~\bibnamefont
  {Hinkley}}, \bibinfo {author} {\bibfnamefont {J.~A.}\ \bibnamefont
  {Sherman}}, \bibinfo {author} {\bibfnamefont {N.~B.}\ \bibnamefont
  {Phillips}}, \bibinfo {author} {\bibfnamefont {M.}~\bibnamefont {Schioppo}},
  \bibinfo {author} {\bibfnamefont {N.~D.}\ \bibnamefont {Lemke}}, \bibinfo
  {author} {\bibfnamefont {K.}~\bibnamefont {Beloy}}, \bibinfo {author}
  {\bibfnamefont {M.}~\bibnamefont {Pizzocaro}}, \bibinfo {author}
  {\bibfnamefont {C.~W.}\ \bibnamefont {Oates}},\ and\ \bibinfo {author}
  {\bibfnamefont {A.~D.}\ \bibnamefont {Ludlow}},\ }\href@noop {} {\bibfield
  {journal} {\bibinfo  {journal} {Science}\ }\textbf {\bibinfo {volume}
  {341}},\ \bibinfo {pages} {1215} (\bibinfo {year} {2013})}\BibitemShut
  {NoStop}%
\bibitem [{\citenamefont {Bloom}\ \emph {et~al.}(2014)\citenamefont {Bloom},
  \citenamefont {Nicholson}, \citenamefont {Williams}, \citenamefont
  {Campbell}, \citenamefont {Bishof}, \citenamefont {Zhang}, \citenamefont
  {Zhang}, \citenamefont {Bromley},\ and\ \citenamefont {Ye}}]{bloom2014clock}%
  \BibitemOpen
  \bibfield  {author} {\bibinfo {author} {\bibfnamefont {B.}~\bibnamefont
  {Bloom}}, \bibinfo {author} {\bibfnamefont {T.}~\bibnamefont {Nicholson}},
  \bibinfo {author} {\bibfnamefont {J.}~\bibnamefont {Williams}}, \bibinfo
  {author} {\bibfnamefont {S.}~\bibnamefont {Campbell}}, \bibinfo {author}
  {\bibfnamefont {M.}~\bibnamefont {Bishof}}, \bibinfo {author} {\bibfnamefont
  {X.}~\bibnamefont {Zhang}}, \bibinfo {author} {\bibfnamefont
  {W.}~\bibnamefont {Zhang}}, \bibinfo {author} {\bibfnamefont
  {S.}~\bibnamefont {Bromley}},\ and\ \bibinfo {author} {\bibfnamefont
  {J.}~\bibnamefont {Ye}},\ }\href@noop {} {\bibfield  {journal} {\bibinfo
  {journal} {Nature}\ }\textbf {\bibinfo {volume} {506}},\ \bibinfo {pages}
  {71} (\bibinfo {year} {2014})}\BibitemShut {NoStop}%
\bibitem [{\citenamefont {Rudolph}\ \emph {et~al.}(2020)\citenamefont
  {Rudolph}, \citenamefont {Wilkason}, \citenamefont {Nantel}, \citenamefont
  {Swan}, \citenamefont {Holland}, \citenamefont {Jiang}, \citenamefont
  {Garber}, \citenamefont {Carman},\ and\ \citenamefont
  {Hogan}}]{rudolf2020_689LMT}%
  \BibitemOpen
  \bibfield  {author} {\bibinfo {author} {\bibfnamefont {J.}~\bibnamefont
  {Rudolph}}, \bibinfo {author} {\bibfnamefont {T.}~\bibnamefont {Wilkason}},
  \bibinfo {author} {\bibfnamefont {M.}~\bibnamefont {Nantel}}, \bibinfo
  {author} {\bibfnamefont {H.}~\bibnamefont {Swan}}, \bibinfo {author}
  {\bibfnamefont {C.~M.}\ \bibnamefont {Holland}}, \bibinfo {author}
  {\bibfnamefont {Y.}~\bibnamefont {Jiang}}, \bibinfo {author} {\bibfnamefont
  {B.~E.}\ \bibnamefont {Garber}}, \bibinfo {author} {\bibfnamefont {S.~P.}\
  \bibnamefont {Carman}},\ and\ \bibinfo {author} {\bibfnamefont {J.~M.}\
  \bibnamefont {Hogan}},\ }\href
  {https://doi.org/10.1103/PhysRevLett.124.083604} {\bibfield  {journal}
  {\bibinfo  {journal} {Phys. Rev. Lett.}\ }\textbf {\bibinfo {volume} {124}},\
  \bibinfo {pages} {083604} (\bibinfo {year} {2020})}\BibitemShut {NoStop}%
\bibitem [{Note1()}]{Note1}%
  \BibitemOpen
  \bibinfo {note} {For atom interferometers based on two-photon atom optics,
  multiple baselines can be used to achieve improved laser noise suppression
  \cite {canuel2018exploring}}\BibitemShut {NoStop}%
\bibitem [{\citenamefont {Taichenachev}\ \emph {et~al.}(2006)\citenamefont
  {Taichenachev}, \citenamefont {Yudin}, \citenamefont {Oates}, \citenamefont
  {Hoyt}, \citenamefont {Barber},\ and\ \citenamefont
  {Hollberg}}]{intercombination_line_magnetig_field}%
  \BibitemOpen
  \bibfield  {author} {\bibinfo {author} {\bibfnamefont {A.~V.}\ \bibnamefont
  {Taichenachev}}, \bibinfo {author} {\bibfnamefont {V.~I.}\ \bibnamefont
  {Yudin}}, \bibinfo {author} {\bibfnamefont {C.~W.}\ \bibnamefont {Oates}},
  \bibinfo {author} {\bibfnamefont {C.~W.}\ \bibnamefont {Hoyt}}, \bibinfo
  {author} {\bibfnamefont {Z.~W.}\ \bibnamefont {Barber}},\ and\ \bibinfo
  {author} {\bibfnamefont {L.}~\bibnamefont {Hollberg}},\ }\href
  {https://doi.org/10.1103/PhysRevLett.96.083001} {\bibfield  {journal}
  {\bibinfo  {journal} {Phys. Rev. Lett.}\ }\textbf {\bibinfo {volume} {96}},\
  \bibinfo {pages} {083001} (\bibinfo {year} {2006})}\BibitemShut {NoStop}%
\bibitem [{\citenamefont {Hu}\ \emph {et~al.}(2017)\citenamefont {Hu},
  \citenamefont {Poli}, \citenamefont {Salvi},\ and\ \citenamefont
  {Tino}}]{Hu2017_clockinterferometry}%
  \BibitemOpen
  \bibfield  {author} {\bibinfo {author} {\bibfnamefont {L.}~\bibnamefont
  {Hu}}, \bibinfo {author} {\bibfnamefont {N.}~\bibnamefont {Poli}}, \bibinfo
  {author} {\bibfnamefont {L.}~\bibnamefont {Salvi}},\ and\ \bibinfo {author}
  {\bibfnamefont {G.~M.}\ \bibnamefont {Tino}},\ }\href
  {https://doi.org/10.1103/PhysRevLett.119.263601} {\bibfield  {journal}
  {\bibinfo  {journal} {Phys. Rev. Lett.}\ }\textbf {\bibinfo {volume} {119}},\
  \bibinfo {pages} {263601} (\bibinfo {year} {2017})}\BibitemShut {NoStop}%
\bibitem [{\citenamefont {Hu}\ \emph {et~al.}(2019)\citenamefont {Hu},
  \citenamefont {Wang}, \citenamefont {Salvi}, \citenamefont {Tinsley},
  \citenamefont {Tino},\ and\ \citenamefont {Poli}}]{hu2019sr}%
  \BibitemOpen
  \bibfield  {author} {\bibinfo {author} {\bibfnamefont {L.}~\bibnamefont
  {Hu}}, \bibinfo {author} {\bibfnamefont {E.}~\bibnamefont {Wang}}, \bibinfo
  {author} {\bibfnamefont {L.}~\bibnamefont {Salvi}}, \bibinfo {author}
  {\bibfnamefont {J.~N.}\ \bibnamefont {Tinsley}}, \bibinfo {author}
  {\bibfnamefont {G.~M.}\ \bibnamefont {Tino}},\ and\ \bibinfo {author}
  {\bibfnamefont {N.}~\bibnamefont {Poli}},\ }\href@noop {} {\bibfield
  {journal} {\bibinfo  {journal} {Classical Quantum Gravity}\ }\textbf
  {\bibinfo {volume} {37}},\ \bibinfo {pages} {014001} (\bibinfo {year}
  {2019})}\BibitemShut {NoStop}%
\bibitem [{\citenamefont {Wilkason}\ \emph {et~al.}(2022)\citenamefont
  {Wilkason}, \citenamefont {Nantel}, \citenamefont {Rudolph}, \citenamefont
  {Jiang}, \citenamefont {Garber}, \citenamefont {Swan}, \citenamefont
  {Carman}, \citenamefont {Abe},\ and\ \citenamefont
  {Hogan}}]{wilkasonFloquet}%
  \BibitemOpen
  \bibfield  {author} {\bibinfo {author} {\bibfnamefont {T.}~\bibnamefont
  {Wilkason}}, \bibinfo {author} {\bibfnamefont {M.}~\bibnamefont {Nantel}},
  \bibinfo {author} {\bibfnamefont {J.}~\bibnamefont {Rudolph}}, \bibinfo
  {author} {\bibfnamefont {Y.}~\bibnamefont {Jiang}}, \bibinfo {author}
  {\bibfnamefont {B.~E.}\ \bibnamefont {Garber}}, \bibinfo {author}
  {\bibfnamefont {H.}~\bibnamefont {Swan}}, \bibinfo {author} {\bibfnamefont
  {S.~P.}\ \bibnamefont {Carman}}, \bibinfo {author} {\bibfnamefont
  {M.}~\bibnamefont {Abe}},\ and\ \bibinfo {author} {\bibfnamefont {J.~M.}\
  \bibnamefont {Hogan}},\ }\href@noop {} {\bibfield  {journal} {\bibinfo
  {journal} {Phys. Rev. Lett.}\ }\textbf {\bibinfo {volume} {129}},\ \bibinfo
  {pages} {183202} (\bibinfo {year} {2022})}\BibitemShut {NoStop}%
\bibitem [{\citenamefont {Arvanitaki}\ \emph {et~al.}(2008)\citenamefont
  {Arvanitaki}, \citenamefont {Dimopoulos}, \citenamefont {Geraci},
  \citenamefont {Hogan},\ and\ \citenamefont
  {Kasevich}}]{Arvanitaki2008_neutrality}%
  \BibitemOpen
  \bibfield  {author} {\bibinfo {author} {\bibfnamefont {A.}~\bibnamefont
  {Arvanitaki}}, \bibinfo {author} {\bibfnamefont {S.}~\bibnamefont
  {Dimopoulos}}, \bibinfo {author} {\bibfnamefont {A.~A.}\ \bibnamefont
  {Geraci}}, \bibinfo {author} {\bibfnamefont {J.}~\bibnamefont {Hogan}},\ and\
  \bibinfo {author} {\bibfnamefont {M.}~\bibnamefont {Kasevich}},\ }\href
  {https://doi.org/10.1103/PhysRevLett.100.120407} {\bibfield  {journal}
  {\bibinfo  {journal} {Phys. Rev. Lett.}\ }\textbf {\bibinfo {volume} {100}},\
  \bibinfo {pages} {120407} (\bibinfo {year} {2008})}\BibitemShut {NoStop}%
\bibitem [{\citenamefont {Levitt}(1986)}]{levitt1986composite}%
  \BibitemOpen
  \bibfield  {author} {\bibinfo {author} {\bibfnamefont {M.~H.}\ \bibnamefont
  {Levitt}},\ }\href@noop {} {\bibfield  {journal} {\bibinfo  {journal}
  {Progress in Nuclear Magnetic Resonance Spectroscopy}\ }\textbf {\bibinfo
  {volume} {18}},\ \bibinfo {pages} {61} (\bibinfo {year} {1986})}\BibitemShut
  {NoStop}%
\bibitem [{\citenamefont {Emsley}\ and\ \citenamefont
  {Bodenhausen}(1992)}]{emsley1992optimization}%
  \BibitemOpen
  \bibfield  {author} {\bibinfo {author} {\bibfnamefont {L.}~\bibnamefont
  {Emsley}}\ and\ \bibinfo {author} {\bibfnamefont {G.}~\bibnamefont
  {Bodenhausen}},\ }\href@noop {} {\bibfield  {journal} {\bibinfo  {journal} {J
  Magn. Reson.}\ }\textbf {\bibinfo {volume} {97}},\ \bibinfo {pages} {135}
  (\bibinfo {year} {1992})}\BibitemShut {NoStop}%
\bibitem [{\citenamefont {Vandersypen}\ and\ \citenamefont
  {Chuang}(2005)}]{NMRComputation}%
  \BibitemOpen
  \bibfield  {author} {\bibinfo {author} {\bibfnamefont {L.~M.~K.}\
  \bibnamefont {Vandersypen}}\ and\ \bibinfo {author} {\bibfnamefont {I.~L.}\
  \bibnamefont {Chuang}},\ }\href {https://doi.org/10.1103/RevModPhys.76.1037}
  {\bibfield  {journal} {\bibinfo  {journal} {Rev. Mod. Phys.}\ }\textbf
  {\bibinfo {volume} {76}},\ \bibinfo {pages} {1037} (\bibinfo {year}
  {2005})}\BibitemShut {NoStop}%
\bibitem [{\citenamefont {Cummins}\ \emph {et~al.}(2003)\citenamefont
  {Cummins}, \citenamefont {Llewellyn},\ and\ \citenamefont
  {Jones}}]{Cummins2003compositecompute}%
  \BibitemOpen
  \bibfield  {author} {\bibinfo {author} {\bibfnamefont {H.~K.}\ \bibnamefont
  {Cummins}}, \bibinfo {author} {\bibfnamefont {G.}~\bibnamefont {Llewellyn}},\
  and\ \bibinfo {author} {\bibfnamefont {J.~A.}\ \bibnamefont {Jones}},\ }\href
  {https://doi.org/10.1103/PhysRevA.67.042308} {\bibfield  {journal} {\bibinfo
  {journal} {Phys. Rev. A}\ }\textbf {\bibinfo {volume} {67}},\ \bibinfo
  {pages} {042308} (\bibinfo {year} {2003})}\BibitemShut {NoStop}%
\bibitem [{\citenamefont {Collin}\ \emph {et~al.}(2004)\citenamefont {Collin},
  \citenamefont {Ithier}, \citenamefont {Aassime}, \citenamefont {Joyez},
  \citenamefont {Vion},\ and\ \citenamefont {Esteve}}]{Collin2004compute}%
  \BibitemOpen
  \bibfield  {author} {\bibinfo {author} {\bibfnamefont {E.}~\bibnamefont
  {Collin}}, \bibinfo {author} {\bibfnamefont {G.}~\bibnamefont {Ithier}},
  \bibinfo {author} {\bibfnamefont {A.}~\bibnamefont {Aassime}}, \bibinfo
  {author} {\bibfnamefont {P.}~\bibnamefont {Joyez}}, \bibinfo {author}
  {\bibfnamefont {D.}~\bibnamefont {Vion}},\ and\ \bibinfo {author}
  {\bibfnamefont {D.}~\bibnamefont {Esteve}},\ }\href
  {https://doi.org/10.1103/PhysRevLett.93.157005} {\bibfield  {journal}
  {\bibinfo  {journal} {Phys. Rev. Lett.}\ }\textbf {\bibinfo {volume} {93}},\
  \bibinfo {pages} {157005} (\bibinfo {year} {2004})}\BibitemShut {NoStop}%
\bibitem [{\citenamefont {Khaneja}\ \emph {et~al.}(2005)\citenamefont
  {Khaneja}, \citenamefont {Reiss}, \citenamefont {Kehlet}, \citenamefont
  {Schulte-Herbr{\"u}ggen},\ and\ \citenamefont {Glaser}}]{khaneja2005optimal}%
  \BibitemOpen
  \bibfield  {author} {\bibinfo {author} {\bibfnamefont {N.}~\bibnamefont
  {Khaneja}}, \bibinfo {author} {\bibfnamefont {T.}~\bibnamefont {Reiss}},
  \bibinfo {author} {\bibfnamefont {C.}~\bibnamefont {Kehlet}}, \bibinfo
  {author} {\bibfnamefont {T.}~\bibnamefont {Schulte-Herbr{\"u}ggen}},\ and\
  \bibinfo {author} {\bibfnamefont {S.~J.}\ \bibnamefont {Glaser}},\
  }\href@noop {} {\bibfield  {journal} {\bibinfo  {journal} {J Magn. Reson.}\
  }\textbf {\bibinfo {volume} {172}},\ \bibinfo {pages} {296} (\bibinfo {year}
  {2005})}\BibitemShut {NoStop}%
\bibitem [{\citenamefont {Doria}\ \emph {et~al.}(2011)\citenamefont {Doria},
  \citenamefont {Calarco},\ and\ \citenamefont {Montangero}}]{Doria2011_CRAB}%
  \BibitemOpen
  \bibfield  {author} {\bibinfo {author} {\bibfnamefont {P.}~\bibnamefont
  {Doria}}, \bibinfo {author} {\bibfnamefont {T.}~\bibnamefont {Calarco}},\
  and\ \bibinfo {author} {\bibfnamefont {S.}~\bibnamefont {Montangero}},\
  }\href {https://doi.org/10.1103/PhysRevLett.106.190501} {\bibfield  {journal}
  {\bibinfo  {journal} {Phys. Rev. Lett.}\ }\textbf {\bibinfo {volume} {106}},\
  \bibinfo {pages} {190501} (\bibinfo {year} {2011})}\BibitemShut {NoStop}%
\bibitem [{\citenamefont {Caneva}\ \emph {et~al.}(2011)\citenamefont {Caneva},
  \citenamefont {Calarco},\ and\ \citenamefont {Montangero}}]{Caneva2011_CRAB}%
  \BibitemOpen
  \bibfield  {author} {\bibinfo {author} {\bibfnamefont {T.}~\bibnamefont
  {Caneva}}, \bibinfo {author} {\bibfnamefont {T.}~\bibnamefont {Calarco}},\
  and\ \bibinfo {author} {\bibfnamefont {S.}~\bibnamefont {Montangero}},\
  }\href {https://doi.org/10.1103/PhysRevA.84.022326} {\bibfield  {journal}
  {\bibinfo  {journal} {Phys. Rev. A}\ }\textbf {\bibinfo {volume} {84}},\
  \bibinfo {pages} {022326} (\bibinfo {year} {2011})}\BibitemShut {NoStop}%
\bibitem [{\citenamefont {Mueller}\ \emph {et~al.}(2022)\citenamefont
  {Mueller}, \citenamefont {Said}, \citenamefont {Jelezko}, \citenamefont
  {Calarco},\ and\ \citenamefont {Montangero}}]{mueller2022one}%
  \BibitemOpen
  \bibfield  {author} {\bibinfo {author} {\bibfnamefont {M.}~\bibnamefont
  {Mueller}}, \bibinfo {author} {\bibfnamefont {R.~S.}\ \bibnamefont {Said}},
  \bibinfo {author} {\bibfnamefont {F.}~\bibnamefont {Jelezko}}, \bibinfo
  {author} {\bibfnamefont {T.}~\bibnamefont {Calarco}},\ and\ \bibinfo {author}
  {\bibfnamefont {S.}~\bibnamefont {Montangero}},\ }\href@noop {} {\bibfield
  {journal} {\bibinfo  {journal} {Rep. Prog. Phys.}\ }\textbf {\bibinfo
  {volume} {85}},\ \bibinfo {pages} {076001} (\bibinfo {year}
  {2022})}\BibitemShut {NoStop}%
\bibitem [{\citenamefont {Grace}\ \emph {et~al.}(2007)\citenamefont {Grace},
  \citenamefont {Brif}, \citenamefont {Rabitz}, \citenamefont {Walmsley},
  \citenamefont {Kosut},\ and\ \citenamefont {Lidar}}]{grace2007optimal}%
  \BibitemOpen
  \bibfield  {author} {\bibinfo {author} {\bibfnamefont {M.}~\bibnamefont
  {Grace}}, \bibinfo {author} {\bibfnamefont {C.}~\bibnamefont {Brif}},
  \bibinfo {author} {\bibfnamefont {H.}~\bibnamefont {Rabitz}}, \bibinfo
  {author} {\bibfnamefont {I.~A.}\ \bibnamefont {Walmsley}}, \bibinfo {author}
  {\bibfnamefont {R.~L.}\ \bibnamefont {Kosut}},\ and\ \bibinfo {author}
  {\bibfnamefont {D.~A.}\ \bibnamefont {Lidar}},\ }\href@noop {} {\bibfield
  {journal} {\bibinfo  {journal} {J Phys. B}\ }\textbf {\bibinfo {volume}
  {40}},\ \bibinfo {pages} {S103} (\bibinfo {year} {2007})}\BibitemShut
  {NoStop}%
\bibitem [{\citenamefont {Rebentrost}\ \emph {et~al.}(2009)\citenamefont
  {Rebentrost}, \citenamefont {Serban}, \citenamefont {Schulte-Herbr\"uggen},\
  and\ \citenamefont {Wilhelm}}]{Rebentrost2009optimal}%
  \BibitemOpen
  \bibfield  {author} {\bibinfo {author} {\bibfnamefont {P.}~\bibnamefont
  {Rebentrost}}, \bibinfo {author} {\bibfnamefont {I.}~\bibnamefont {Serban}},
  \bibinfo {author} {\bibfnamefont {T.}~\bibnamefont {Schulte-Herbr\"uggen}},\
  and\ \bibinfo {author} {\bibfnamefont {F.~K.}\ \bibnamefont {Wilhelm}},\
  }\href {https://doi.org/10.1103/PhysRevLett.102.090401} {\bibfield  {journal}
  {\bibinfo  {journal} {Phys. Rev. Lett.}\ }\textbf {\bibinfo {volume} {102}},\
  \bibinfo {pages} {090401} (\bibinfo {year} {2009})}\BibitemShut {NoStop}%
\bibitem [{\citenamefont {Abdelhafez}\ \emph {et~al.}(2020)\citenamefont
  {Abdelhafez}, \citenamefont {Baker}, \citenamefont {Gyenis}, \citenamefont
  {Mundada}, \citenamefont {Houck}, \citenamefont {Schuster},\ and\
  \citenamefont {Koch}}]{Abdelhafez2020optimal}%
  \BibitemOpen
  \bibfield  {author} {\bibinfo {author} {\bibfnamefont {M.}~\bibnamefont
  {Abdelhafez}}, \bibinfo {author} {\bibfnamefont {B.}~\bibnamefont {Baker}},
  \bibinfo {author} {\bibfnamefont {A.}~\bibnamefont {Gyenis}}, \bibinfo
  {author} {\bibfnamefont {P.}~\bibnamefont {Mundada}}, \bibinfo {author}
  {\bibfnamefont {A.~A.}\ \bibnamefont {Houck}}, \bibinfo {author}
  {\bibfnamefont {D.}~\bibnamefont {Schuster}},\ and\ \bibinfo {author}
  {\bibfnamefont {J.}~\bibnamefont {Koch}},\ }\href
  {https://doi.org/10.1103/PhysRevA.101.022321} {\bibfield  {journal} {\bibinfo
   {journal} {Phys. Rev. A}\ }\textbf {\bibinfo {volume} {101}},\ \bibinfo
  {pages} {022321} (\bibinfo {year} {2020})}\BibitemShut {NoStop}%
\bibitem [{\citenamefont {Koch}\ \emph {et~al.}(2022)\citenamefont {Koch},
  \citenamefont {Boscain}, \citenamefont {Calarco}, \citenamefont {Dirr},
  \citenamefont {Filipp}, \citenamefont {Glaser}, \citenamefont {Kosloff},
  \citenamefont {Montangero}, \citenamefont {Schulte-Herbr{\"u}ggen},
  \citenamefont {Sugny} \emph {et~al.}}]{koch2022quantum}%
  \BibitemOpen
  \bibfield  {author} {\bibinfo {author} {\bibfnamefont {C.~P.}\ \bibnamefont
  {Koch}}, \bibinfo {author} {\bibfnamefont {U.}~\bibnamefont {Boscain}},
  \bibinfo {author} {\bibfnamefont {T.}~\bibnamefont {Calarco}}, \bibinfo
  {author} {\bibfnamefont {G.}~\bibnamefont {Dirr}}, \bibinfo {author}
  {\bibfnamefont {S.}~\bibnamefont {Filipp}}, \bibinfo {author} {\bibfnamefont
  {S.~J.}\ \bibnamefont {Glaser}}, \bibinfo {author} {\bibfnamefont
  {R.}~\bibnamefont {Kosloff}}, \bibinfo {author} {\bibfnamefont
  {S.}~\bibnamefont {Montangero}}, \bibinfo {author} {\bibfnamefont
  {T.}~\bibnamefont {Schulte-Herbr{\"u}ggen}}, \bibinfo {author} {\bibfnamefont
  {D.}~\bibnamefont {Sugny}}, \emph {et~al.},\ }\href@noop {} {\bibfield
  {journal} {\bibinfo  {journal} {EPJ Quantum Technol.}\ }\textbf {\bibinfo
  {volume} {9}},\ \bibinfo {pages} {19} (\bibinfo {year} {2022})}\BibitemShut
  {NoStop}%
\bibitem [{\citenamefont {Magrini}\ \emph {et~al.}(2021)\citenamefont
  {Magrini}, \citenamefont {Rosenzweig}, \citenamefont {Bach}, \citenamefont
  {Deutschmann-Olek}, \citenamefont {Hofer}, \citenamefont {Hong},
  \citenamefont {Kiesel}, \citenamefont {Kugi},\ and\ \citenamefont
  {Aspelmeyer}}]{magrini2021real}%
  \BibitemOpen
  \bibfield  {author} {\bibinfo {author} {\bibfnamefont {L.}~\bibnamefont
  {Magrini}}, \bibinfo {author} {\bibfnamefont {P.}~\bibnamefont {Rosenzweig}},
  \bibinfo {author} {\bibfnamefont {C.}~\bibnamefont {Bach}}, \bibinfo {author}
  {\bibfnamefont {A.}~\bibnamefont {Deutschmann-Olek}}, \bibinfo {author}
  {\bibfnamefont {S.~G.}\ \bibnamefont {Hofer}}, \bibinfo {author}
  {\bibfnamefont {S.}~\bibnamefont {Hong}}, \bibinfo {author} {\bibfnamefont
  {N.}~\bibnamefont {Kiesel}}, \bibinfo {author} {\bibfnamefont
  {A.}~\bibnamefont {Kugi}},\ and\ \bibinfo {author} {\bibfnamefont
  {M.}~\bibnamefont {Aspelmeyer}},\ }\href@noop {} {\bibfield  {journal}
  {\bibinfo  {journal} {Nature}\ }\textbf {\bibinfo {volume} {595}},\ \bibinfo
  {pages} {373} (\bibinfo {year} {2021})}\BibitemShut {NoStop}%
\bibitem [{\citenamefont {Weiss}\ \emph {et~al.}(2021)\citenamefont {Weiss},
  \citenamefont {Roda-Llordes}, \citenamefont {Torrontegui}, \citenamefont
  {Aspelmeyer},\ and\ \citenamefont {Romero-Isart}}]{weiss2021large}%
  \BibitemOpen
  \bibfield  {author} {\bibinfo {author} {\bibfnamefont {T.}~\bibnamefont
  {Weiss}}, \bibinfo {author} {\bibfnamefont {M.}~\bibnamefont {Roda-Llordes}},
  \bibinfo {author} {\bibfnamefont {E.}~\bibnamefont {Torrontegui}}, \bibinfo
  {author} {\bibfnamefont {M.}~\bibnamefont {Aspelmeyer}},\ and\ \bibinfo
  {author} {\bibfnamefont {O.}~\bibnamefont {Romero-Isart}},\ }\href@noop {}
  {\bibfield  {journal} {\bibinfo  {journal} {Phys. Rev. Lett.}\ }\textbf
  {\bibinfo {volume} {127}},\ \bibinfo {pages} {023601} (\bibinfo {year}
  {2021})}\BibitemShut {NoStop}%
\bibitem [{\citenamefont {Butts}\ \emph {et~al.}(2013)\citenamefont {Butts},
  \citenamefont {Kotru}, \citenamefont {Kinast}, \citenamefont {Radojevic},
  \citenamefont {Timmons},\ and\ \citenamefont {Stoner}}]{Butts2013}%
  \BibitemOpen
  \bibfield  {author} {\bibinfo {author} {\bibfnamefont {D.~L.}\ \bibnamefont
  {Butts}}, \bibinfo {author} {\bibfnamefont {K.}~\bibnamefont {Kotru}},
  \bibinfo {author} {\bibfnamefont {J.~M.}\ \bibnamefont {Kinast}}, \bibinfo
  {author} {\bibfnamefont {A.~M.}\ \bibnamefont {Radojevic}}, \bibinfo {author}
  {\bibfnamefont {B.~P.}\ \bibnamefont {Timmons}},\ and\ \bibinfo {author}
  {\bibfnamefont {R.~E.}\ \bibnamefont {Stoner}},\ }\href
  {https://doi.org/10.1364/JOSAB.30.000922} {\bibfield  {journal} {\bibinfo
  {journal} {J. Opt. Soc. Am. B}\ }\textbf {\bibinfo {volume} {30}},\ \bibinfo
  {pages} {922} (\bibinfo {year} {2013})}\BibitemShut {NoStop}%
\bibitem [{\citenamefont {Dunning}\ \emph {et~al.}(2014)\citenamefont
  {Dunning}, \citenamefont {Gregory}, \citenamefont {Bateman}, \citenamefont
  {Cooper}, \citenamefont {Himsworth}, \citenamefont {Jones},\ and\
  \citenamefont {Freegarde}}]{dunning2014_composite}%
  \BibitemOpen
  \bibfield  {author} {\bibinfo {author} {\bibfnamefont {A.}~\bibnamefont
  {Dunning}}, \bibinfo {author} {\bibfnamefont {R.}~\bibnamefont {Gregory}},
  \bibinfo {author} {\bibfnamefont {J.}~\bibnamefont {Bateman}}, \bibinfo
  {author} {\bibfnamefont {N.}~\bibnamefont {Cooper}}, \bibinfo {author}
  {\bibfnamefont {M.}~\bibnamefont {Himsworth}}, \bibinfo {author}
  {\bibfnamefont {J.~A.}\ \bibnamefont {Jones}},\ and\ \bibinfo {author}
  {\bibfnamefont {T.}~\bibnamefont {Freegarde}},\ }\href
  {https://doi.org/10.1103/PhysRevA.90.033608} {\bibfield  {journal} {\bibinfo
  {journal} {Phys. Rev. A}\ }\textbf {\bibinfo {volume} {90}},\ \bibinfo
  {pages} {033608} (\bibinfo {year} {2014})}\BibitemShut {NoStop}%
\bibitem [{\citenamefont {Berg}\ \emph {et~al.}(2015)\citenamefont {Berg},
  \citenamefont {Abend}, \citenamefont {Tackmann}, \citenamefont {Schubert},
  \citenamefont {Giese}, \citenamefont {Schleich}, \citenamefont {Narducci},
  \citenamefont {Ertmer},\ and\ \citenamefont {Rasel}}]{Berg2015_composite}%
  \BibitemOpen
  \bibfield  {author} {\bibinfo {author} {\bibfnamefont {P.}~\bibnamefont
  {Berg}}, \bibinfo {author} {\bibfnamefont {S.}~\bibnamefont {Abend}},
  \bibinfo {author} {\bibfnamefont {G.}~\bibnamefont {Tackmann}}, \bibinfo
  {author} {\bibfnamefont {C.}~\bibnamefont {Schubert}}, \bibinfo {author}
  {\bibfnamefont {E.}~\bibnamefont {Giese}}, \bibinfo {author} {\bibfnamefont
  {W.~P.}\ \bibnamefont {Schleich}}, \bibinfo {author} {\bibfnamefont {F.~A.}\
  \bibnamefont {Narducci}}, \bibinfo {author} {\bibfnamefont {W.}~\bibnamefont
  {Ertmer}},\ and\ \bibinfo {author} {\bibfnamefont {E.~M.}\ \bibnamefont
  {Rasel}},\ }\href {https://doi.org/10.1103/PhysRevLett.114.063002} {\bibfield
   {journal} {\bibinfo  {journal} {Phys. Rev. Lett.}\ }\textbf {\bibinfo
  {volume} {114}},\ \bibinfo {pages} {063002} (\bibinfo {year}
  {2015})}\BibitemShut {NoStop}%
\bibitem [{\citenamefont {Luo}\ \emph {et~al.}(2016)\citenamefont {Luo},
  \citenamefont {Yan}, \citenamefont {Hu}, \citenamefont {Jia}, \citenamefont
  {Wei},\ and\ \citenamefont {Yang}}]{Luo2016_shaped_pulses}%
  \BibitemOpen
  \bibfield  {author} {\bibinfo {author} {\bibfnamefont {Y.}~\bibnamefont
  {Luo}}, \bibinfo {author} {\bibfnamefont {S.}~\bibnamefont {Yan}}, \bibinfo
  {author} {\bibfnamefont {Q.}~\bibnamefont {Hu}}, \bibinfo {author}
  {\bibfnamefont {A.}~\bibnamefont {Jia}}, \bibinfo {author} {\bibfnamefont
  {C.}~\bibnamefont {Wei}},\ and\ \bibinfo {author} {\bibfnamefont
  {J.}~\bibnamefont {Yang}},\ }\href@noop {} {\bibfield  {journal} {\bibinfo
  {journal} {Eur. Phys. J. D}\ }\textbf {\bibinfo {volume} {70}} (\bibinfo
  {year} {2016})}\BibitemShut {NoStop}%
\bibitem [{\citenamefont {Kovachy}\ \emph {et~al.}(2012)\citenamefont
  {Kovachy}, \citenamefont {Chiow},\ and\ \citenamefont
  {Kasevich}}]{tim_ARP_Bragg}%
  \BibitemOpen
  \bibfield  {author} {\bibinfo {author} {\bibfnamefont {T.}~\bibnamefont
  {Kovachy}}, \bibinfo {author} {\bibfnamefont {S.-w.}\ \bibnamefont {Chiow}},\
  and\ \bibinfo {author} {\bibfnamefont {M.~A.}\ \bibnamefont {Kasevich}},\
  }\href {https://doi.org/10.1103/PhysRevA.86.011606} {\bibfield  {journal}
  {\bibinfo  {journal} {Phys. Rev. A}\ }\textbf {\bibinfo {volume} {86}},\
  \bibinfo {pages} {011606} (\bibinfo {year} {2012})}\BibitemShut {NoStop}%
\bibitem [{\citenamefont {Saywell}\ \emph {et~al.}(2018)\citenamefont
  {Saywell}, \citenamefont {Kuprov}, \citenamefont {Goodwin}, \citenamefont
  {Carey},\ and\ \citenamefont {Freegarde}}]{saywell2018_QuOC_Mirror}%
  \BibitemOpen
  \bibfield  {author} {\bibinfo {author} {\bibfnamefont {J.~C.}\ \bibnamefont
  {Saywell}}, \bibinfo {author} {\bibfnamefont {I.}~\bibnamefont {Kuprov}},
  \bibinfo {author} {\bibfnamefont {D.}~\bibnamefont {Goodwin}}, \bibinfo
  {author} {\bibfnamefont {M.}~\bibnamefont {Carey}},\ and\ \bibinfo {author}
  {\bibfnamefont {T.}~\bibnamefont {Freegarde}},\ }\href
  {https://doi.org/10.1103/PhysRevA.98.023625} {\bibfield  {journal} {\bibinfo
  {journal} {Phys. Rev. A}\ }\textbf {\bibinfo {volume} {98}},\ \bibinfo
  {pages} {023625} (\bibinfo {year} {2018})}\BibitemShut {NoStop}%
\bibitem [{\citenamefont {Saywell}\ \emph
  {et~al.}(2020{\natexlab{a}})\citenamefont {Saywell}, \citenamefont {Carey},
  \citenamefont {Belal}, \citenamefont {Kuprov},\ and\ \citenamefont
  {Freegarde}}]{saywell2020optimal}%
  \BibitemOpen
  \bibfield  {author} {\bibinfo {author} {\bibfnamefont {J.}~\bibnamefont
  {Saywell}}, \bibinfo {author} {\bibfnamefont {M.}~\bibnamefont {Carey}},
  \bibinfo {author} {\bibfnamefont {M.}~\bibnamefont {Belal}}, \bibinfo
  {author} {\bibfnamefont {I.}~\bibnamefont {Kuprov}},\ and\ \bibinfo {author}
  {\bibfnamefont {T.}~\bibnamefont {Freegarde}},\ }\href@noop {} {\bibfield
  {journal} {\bibinfo  {journal} {J Phys. B}\ }\textbf {\bibinfo {volume}
  {53}},\ \bibinfo {pages} {085006} (\bibinfo {year}
  {2020}{\natexlab{a}})}\BibitemShut {NoStop}%
\bibitem [{\citenamefont {Saywell}\ \emph
  {et~al.}(2020{\natexlab{b}})\citenamefont {Saywell}, \citenamefont {Carey},
  \citenamefont {Kuprov},\ and\ \citenamefont
  {Freegarde}}]{saywell_2020_biselective}%
  \BibitemOpen
  \bibfield  {author} {\bibinfo {author} {\bibfnamefont {J.}~\bibnamefont
  {Saywell}}, \bibinfo {author} {\bibfnamefont {M.}~\bibnamefont {Carey}},
  \bibinfo {author} {\bibfnamefont {I.}~\bibnamefont {Kuprov}},\ and\ \bibinfo
  {author} {\bibfnamefont {T.}~\bibnamefont {Freegarde}},\ }\href
  {https://doi.org/10.1103/PhysRevA.101.063625} {\bibfield  {journal} {\bibinfo
   {journal} {Phys. Rev. A}\ }\textbf {\bibinfo {volume} {101}},\ \bibinfo
  {pages} {063625} (\bibinfo {year} {2020}{\natexlab{b}})}\BibitemShut
  {NoStop}%
\bibitem [{\citenamefont {Goerz}\ \emph {et~al.}(2021)\citenamefont {Goerz},
  \citenamefont {Kasevich},\ and\ \citenamefont
  {Malinovsky}}]{goerz2021quantum}%
  \BibitemOpen
  \bibfield  {author} {\bibinfo {author} {\bibfnamefont {M.~H.}\ \bibnamefont
  {Goerz}}, \bibinfo {author} {\bibfnamefont {M.~A.}\ \bibnamefont
  {Kasevich}},\ and\ \bibinfo {author} {\bibfnamefont {V.~S.}\ \bibnamefont
  {Malinovsky}},\ }in\ \href@noop {} {\emph {\bibinfo {booktitle} {Optical and
  Quantum Sensing and Precision Metrology}}},\ Vol.\ \bibinfo {volume} {11700}\
  (\bibinfo {organization} {SPIE},\ \bibinfo {year} {2021})\ p.\ \bibinfo
  {pages} {1170005}\BibitemShut {NoStop}%
\bibitem [{\citenamefont {Goerz}\ \emph {et~al.}(2023)\citenamefont {Goerz},
  \citenamefont {Kasevich},\ and\ \citenamefont {Malinovsky}}]{Goerz2023}%
  \BibitemOpen
  \bibfield  {author} {\bibinfo {author} {\bibfnamefont {M.~H.}\ \bibnamefont
  {Goerz}}, \bibinfo {author} {\bibfnamefont {M.~A.}\ \bibnamefont
  {Kasevich}},\ and\ \bibinfo {author} {\bibfnamefont {V.~S.}\ \bibnamefont
  {Malinovsky}},\ }\href {https://doi.org/10.3390/atoms11020036} {\bibfield
  {journal} {\bibinfo  {journal} {Atoms}\ }\textbf {\bibinfo {volume} {11}},\
  \bibinfo {pages} {36} (\bibinfo {year} {2023})}\BibitemShut {NoStop}%
\bibitem [{\citenamefont {Ball}\ \emph {et~al.}(2021)\citenamefont {Ball},
  \citenamefont {Biercuk}, \citenamefont {Carvalho}, \citenamefont {Chen},
  \citenamefont {Hush}, \citenamefont {De~Castro}, \citenamefont {Li},
  \citenamefont {Liebermann}, \citenamefont {Slatyer}, \citenamefont {Edmunds}
  \emph {et~al.}}]{ball2021software}%
  \BibitemOpen
  \bibfield  {author} {\bibinfo {author} {\bibfnamefont {H.}~\bibnamefont
  {Ball}}, \bibinfo {author} {\bibfnamefont {M.~J.}\ \bibnamefont {Biercuk}},
  \bibinfo {author} {\bibfnamefont {A.~R.}\ \bibnamefont {Carvalho}}, \bibinfo
  {author} {\bibfnamefont {J.}~\bibnamefont {Chen}}, \bibinfo {author}
  {\bibfnamefont {M.}~\bibnamefont {Hush}}, \bibinfo {author} {\bibfnamefont
  {L.~A.}\ \bibnamefont {De~Castro}}, \bibinfo {author} {\bibfnamefont
  {L.}~\bibnamefont {Li}}, \bibinfo {author} {\bibfnamefont {P.~J.}\
  \bibnamefont {Liebermann}}, \bibinfo {author} {\bibfnamefont {H.~J.}\
  \bibnamefont {Slatyer}}, \bibinfo {author} {\bibfnamefont {C.}~\bibnamefont
  {Edmunds}}, \emph {et~al.},\ }\href@noop {} {\bibfield  {journal} {\bibinfo
  {journal} {Quantum Sci. Technol.}\ }\textbf {\bibinfo {volume} {6}},\
  \bibinfo {pages} {044011} (\bibinfo {year} {2021})}\BibitemShut {NoStop}%
\bibitem [{\citenamefont {Bertoldi}\ \emph {et~al.}(2021)\citenamefont
  {Bertoldi}, \citenamefont {Feng}, \citenamefont {Naik}, \citenamefont
  {Canuel}, \citenamefont {Bouyer},\ and\ \citenamefont
  {Prevedelli}}]{bertoldi2021fast}%
  \BibitemOpen
  \bibfield  {author} {\bibinfo {author} {\bibfnamefont {A.}~\bibnamefont
  {Bertoldi}}, \bibinfo {author} {\bibfnamefont {C.-H.}\ \bibnamefont {Feng}},
  \bibinfo {author} {\bibfnamefont {D.}~\bibnamefont {Naik}}, \bibinfo {author}
  {\bibfnamefont {B.}~\bibnamefont {Canuel}}, \bibinfo {author} {\bibfnamefont
  {P.}~\bibnamefont {Bouyer}},\ and\ \bibinfo {author} {\bibfnamefont
  {M.}~\bibnamefont {Prevedelli}},\ }\href@noop {} {\bibfield  {journal}
  {\bibinfo  {journal} {Phys. Rev. Lett.}\ }\textbf {\bibinfo {volume} {127}},\
  \bibinfo {pages} {013202} (\bibinfo {year} {2021})}\BibitemShut {NoStop}%
\bibitem [{\citenamefont {Nourshargh}\ \emph {et~al.}(2021)\citenamefont
  {Nourshargh}, \citenamefont {Lellouch}, \citenamefont {Hedges}, \citenamefont
  {Langlois}, \citenamefont {Bongs},\ and\ \citenamefont
  {Holynski}}]{nourshargh2021circulating}%
  \BibitemOpen
  \bibfield  {author} {\bibinfo {author} {\bibfnamefont {R.}~\bibnamefont
  {Nourshargh}}, \bibinfo {author} {\bibfnamefont {S.}~\bibnamefont
  {Lellouch}}, \bibinfo {author} {\bibfnamefont {S.}~\bibnamefont {Hedges}},
  \bibinfo {author} {\bibfnamefont {M.}~\bibnamefont {Langlois}}, \bibinfo
  {author} {\bibfnamefont {K.}~\bibnamefont {Bongs}},\ and\ \bibinfo {author}
  {\bibfnamefont {M.}~\bibnamefont {Holynski}},\ }\href@noop {} {\bibfield
  {journal} {\bibinfo  {journal} {Comm. Phys.}\ }\textbf {\bibinfo {volume}
  {4}},\ \bibinfo {pages} {1} (\bibinfo {year} {2021})}\BibitemShut {NoStop}%
\bibitem [{\citenamefont {Goerz}\ \emph {et~al.}(2014)\citenamefont {Goerz},
  \citenamefont {Halperin}, \citenamefont {Aytac}, \citenamefont {Koch},\ and\
  \citenamefont {Whaley}}]{goerz2014robustness}%
  \BibitemOpen
  \bibfield  {author} {\bibinfo {author} {\bibfnamefont {M.~H.}\ \bibnamefont
  {Goerz}}, \bibinfo {author} {\bibfnamefont {E.~J.}\ \bibnamefont {Halperin}},
  \bibinfo {author} {\bibfnamefont {J.~M.}\ \bibnamefont {Aytac}}, \bibinfo
  {author} {\bibfnamefont {C.~P.}\ \bibnamefont {Koch}},\ and\ \bibinfo
  {author} {\bibfnamefont {K.~B.}\ \bibnamefont {Whaley}},\ }\href@noop {}
  {\bibfield  {journal} {\bibinfo  {journal} {Phys. Rev. A}\ }\textbf {\bibinfo
  {volume} {90}},\ \bibinfo {pages} {032329} (\bibinfo {year}
  {2014})}\BibitemShut {NoStop}%
\bibitem [{\citenamefont {Genov}\ and\ \citenamefont
  {Vitanov}(2013)}]{genov2013dynamical}%
  \BibitemOpen
  \bibfield  {author} {\bibinfo {author} {\bibfnamefont {G.~T.}\ \bibnamefont
  {Genov}}\ and\ \bibinfo {author} {\bibfnamefont {N.~V.}\ \bibnamefont
  {Vitanov}},\ }\href@noop {} {\bibfield  {journal} {\bibinfo  {journal} {Phys.
  Rev. Lett.}\ }\textbf {\bibinfo {volume} {110}},\ \bibinfo {pages} {133002}
  (\bibinfo {year} {2013})}\BibitemShut {NoStop}%
\bibitem [{\citenamefont {Deppner}\ \emph {et~al.}(2021)\citenamefont
  {Deppner}, \citenamefont {Herr}, \citenamefont {Cornelius}, \citenamefont
  {Stromberger}, \citenamefont {Sternke}, \citenamefont {Grzeschik},
  \citenamefont {Grote}, \citenamefont {Rudolph}, \citenamefont {Herrmann},
  \citenamefont {Krutzik}, \citenamefont {Wenzlawski}, \citenamefont {Corgier},
  \citenamefont {Charron}, \citenamefont {Gu\'ery-Odelin}, \citenamefont
  {Gaaloul}, \citenamefont {L\"ammerzahl}, \citenamefont {Peters},
  \citenamefont {Windpassinger},\ and\ \citenamefont {Rasel}}]{Deppner2021}%
  \BibitemOpen
  \bibfield  {author} {\bibinfo {author} {\bibfnamefont {C.}~\bibnamefont
  {Deppner}}, \bibinfo {author} {\bibfnamefont {W.}~\bibnamefont {Herr}},
  \bibinfo {author} {\bibfnamefont {M.}~\bibnamefont {Cornelius}}, \bibinfo
  {author} {\bibfnamefont {P.}~\bibnamefont {Stromberger}}, \bibinfo {author}
  {\bibfnamefont {T.}~\bibnamefont {Sternke}}, \bibinfo {author} {\bibfnamefont
  {C.}~\bibnamefont {Grzeschik}}, \bibinfo {author} {\bibfnamefont
  {A.}~\bibnamefont {Grote}}, \bibinfo {author} {\bibfnamefont
  {J.}~\bibnamefont {Rudolph}}, \bibinfo {author} {\bibfnamefont
  {S.}~\bibnamefont {Herrmann}}, \bibinfo {author} {\bibfnamefont
  {M.}~\bibnamefont {Krutzik}}, \bibinfo {author} {\bibfnamefont
  {A.}~\bibnamefont {Wenzlawski}}, \bibinfo {author} {\bibfnamefont
  {R.}~\bibnamefont {Corgier}}, \bibinfo {author} {\bibfnamefont
  {E.}~\bibnamefont {Charron}}, \bibinfo {author} {\bibfnamefont
  {D.}~\bibnamefont {Gu\'ery-Odelin}}, \bibinfo {author} {\bibfnamefont
  {N.}~\bibnamefont {Gaaloul}}, \bibinfo {author} {\bibfnamefont
  {C.}~\bibnamefont {L\"ammerzahl}}, \bibinfo {author} {\bibfnamefont
  {A.}~\bibnamefont {Peters}}, \bibinfo {author} {\bibfnamefont
  {P.}~\bibnamefont {Windpassinger}},\ and\ \bibinfo {author} {\bibfnamefont
  {E.~M.}\ \bibnamefont {Rasel}},\ }\href
  {https://doi.org/10.1103/PhysRevLett.127.100401} {\bibfield  {journal}
  {\bibinfo  {journal} {Phys. Rev. Lett.}\ }\textbf {\bibinfo {volume} {127}},\
  \bibinfo {pages} {100401} (\bibinfo {year} {2021})}\BibitemShut {NoStop}%
\bibitem [{\citenamefont {Kovachy}\ \emph
  {et~al.}(2015{\natexlab{b}})\citenamefont {Kovachy}, \citenamefont {Hogan},
  \citenamefont {Sugarbaker}, \citenamefont {Dickerson}, \citenamefont
  {Donnelly}, \citenamefont {Overstreet},\ and\ \citenamefont
  {Kasevich}}]{Kovachy2015Lensing}%
  \BibitemOpen
  \bibfield  {author} {\bibinfo {author} {\bibfnamefont {T.}~\bibnamefont
  {Kovachy}}, \bibinfo {author} {\bibfnamefont {J.~M.}\ \bibnamefont {Hogan}},
  \bibinfo {author} {\bibfnamefont {A.}~\bibnamefont {Sugarbaker}}, \bibinfo
  {author} {\bibfnamefont {S.~M.}\ \bibnamefont {Dickerson}}, \bibinfo {author}
  {\bibfnamefont {C.~A.}\ \bibnamefont {Donnelly}}, \bibinfo {author}
  {\bibfnamefont {C.}~\bibnamefont {Overstreet}},\ and\ \bibinfo {author}
  {\bibfnamefont {M.~A.}\ \bibnamefont {Kasevich}},\ }\href
  {https://doi.org/10.1103/PhysRevLett.114.143004} {\bibfield  {journal}
  {\bibinfo  {journal} {Phys. Rev. Lett.}\ }\textbf {\bibinfo {volume} {114}},\
  \bibinfo {pages} {143004} (\bibinfo {year} {2015}{\natexlab{b}})}\BibitemShut
  {NoStop}%
\bibitem [{Note2()}]{Note2}%
  \BibitemOpen
  \bibinfo {note} {For instance, using commercially available laser systems
  from Menlo Systems}\BibitemShut {NoStop}%
\bibitem [{\citenamefont {Feng}\ \emph {et~al.}(2018)\citenamefont {Feng},
  \citenamefont {Cho}, \citenamefont {Katiyar}, \citenamefont {Li},
  \citenamefont {Lu}, \citenamefont {Baugh},\ and\ \citenamefont
  {Laflamme}}]{Feng2018}%
  \BibitemOpen
  \bibfield  {author} {\bibinfo {author} {\bibfnamefont {G.}~\bibnamefont
  {Feng}}, \bibinfo {author} {\bibfnamefont {F.~H.}\ \bibnamefont {Cho}},
  \bibinfo {author} {\bibfnamefont {H.}~\bibnamefont {Katiyar}}, \bibinfo
  {author} {\bibfnamefont {J.}~\bibnamefont {Li}}, \bibinfo {author}
  {\bibfnamefont {D.}~\bibnamefont {Lu}}, \bibinfo {author} {\bibfnamefont
  {J.}~\bibnamefont {Baugh}},\ and\ \bibinfo {author} {\bibfnamefont
  {R.}~\bibnamefont {Laflamme}},\ }\href
  {https://doi.org/10.1103/PhysRevA.98.052341} {\bibfield  {journal} {\bibinfo
  {journal} {Phys. Rev. A}\ }\textbf {\bibinfo {volume} {98}},\ \bibinfo
  {pages} {052341} (\bibinfo {year} {2018})}\BibitemShut {NoStop}%
\bibitem [{\citenamefont {Boyd}(2007)}]{boydthesis}%
  \BibitemOpen
  \bibfield  {author} {\bibinfo {author} {\bibfnamefont {M.}~\bibnamefont
  {Boyd}},\ }\emph {\bibinfo {title} {High Precision Spectroscopy of Strontium
  in an Optical Lattice: Towards a New Standard for Frequency and Time}},\
  \href@noop {} {Ph.D. thesis},\ \bibinfo  {school} {University of Colorado}
  (\bibinfo {year} {2007})\BibitemShut {NoStop}%
\bibitem [{\citenamefont {Metcalf}\ and\ \citenamefont {Van~der
  Straten}(1999)}]{metcalf1999laser}%
  \BibitemOpen
  \bibfield  {author} {\bibinfo {author} {\bibfnamefont {H.~J.}\ \bibnamefont
  {Metcalf}}\ and\ \bibinfo {author} {\bibfnamefont {P.}~\bibnamefont {Van~der
  Straten}},\ }\href@noop {} {\emph {\bibinfo {title} {Laser cooling and
  trapping}}}\ (\bibinfo  {publisher} {Springer Science \& Business Media},\
  \bibinfo {year} {1999})\BibitemShut {NoStop}%
\end{thebibliography}
\twocolumngrid

\end{document}